\documentclass[lettersize,journal]{IEEEtran}
\usepackage{amsmath,amsfonts}
\usepackage{amssymb}
\usepackage{amsthm}
\usepackage{algorithmic}
\usepackage{algorithm}
\usepackage{array}
\usepackage[caption=false,font=normalsize,labelfont=sf,textfont=sf]{subfig}
\usepackage{textcomp}
\usepackage{stfloats}
\usepackage{url}
\usepackage{verbatim}
\usepackage{graphicx}
\usepackage{cite}
\usepackage{pifont}
\usepackage{booktabs}      
\usepackage{multirow}      
\usepackage{color}   
\usepackage{bm}
\usepackage{balance}
\newcommand{\cmark}{\ding{51}} 
\newcommand{\xmark}{\ding{55}} 
\newtheorem{rmk}{Remark} 
\def \eg {\emph{e.g.}}
\def \ie {\emph{i.e.}}
\def \etal {\emph{et al.}}

\DeclareMathOperator*{\argmin}{arg\,min}

\begin{document}

\title{DUN-SRE: Deep Unrolling Network with Spatiotemporal Rotation Equivariance for Dynamic MRI Reconstruction}

\author{Yuliang Zhu, Jing Cheng, Qi Xie, Zhuo-Xu Cui, Qingyong Zhu, Yuanyuan Liu, Xin Liu, Jianfeng Ren,~\IEEEmembership{Senior Member,~IEEE,} Chengbo Wang, Dong Liang,~\IEEEmembership{Senior Member,~IEEE,}
\thanks{This work was supported in part by the National Natural Science Foundation of China under Grant nos. 62322119, 62201561, 62206273, 62476268, 12226008, 62125111, (\textit{Corresponding author:Dong Liang, Chengbo Wang.)}}
\thanks{Yuliang Zhu is with University of Nottingham Ningbo China, Ningbo, Zhejiang, China, and also with Shenzhen Institute of Advanced Technology, Chinese Academy of Sciences, Shenzhen, Guangdong, China.}
\thanks{Jing Cheng, Zhuo-Xu Cui, Qingyong Zhu, Yuanyuan Liu, Xin Liu, Dong Liang are with Shenzhen Institute of Advanced Technology, Chinese Academy of Sciences, Shenzhen, Guangdong, China.}
\thanks{Qi Xie is with the School of Mathematics and Statistics, Xi’an Jiaotong University, Xi’an, Shaanxi, China.}
\thanks{Jianfeng Ren, Chengbo Wang are with University of Nottingham Ningbo China, Ningbo, Zhejiang, China.}
\thanks{Yuliang Zhu and Jing Cheng contributed equally to this study.}}

\markboth{IEEE Journal of Selected Topics in Signal Processing}%
{ZHU \MakeLowercase{\textit{et al.}}: A Sample Article Using IEEEtran.cls for IEEE Journals}

\IEEEpubid{}

\maketitle

\begin{abstract}
Dynamic Magnetic Resonance Imaging (MRI) exhibits transformation symmetries, including spatial rotation symmetry within individual frames and temporal symmetry along the time dimension. Explicit incorporation of these symmetry priors in the reconstruction model can significantly improve image quality, especially under aggressive undersampling scenarios. Recently, Equivariant convolutional neural network (ECNN) has shown great promise in exploiting spatial symmetry priors. However, existing ECNNs critically fail to model temporal symmetry, arguably the most universal and informative structural prior in dynamic MRI reconstruction. 
To tackle this issue, we propose a novel Deep Unrolling Network with Spatiotemporal Rotation Equivariance (DUN-SRE) for Dynamic MRI Reconstruction. 
The DUN-SRE establishes spatiotemporal equivariance through a (2+1)D equivariant convolutional architecture. In particular, it integrates both the data consistency and proximal mapping 
module into a unified deep unrolling framework. This architecture ensures rigorous propagation of spatiotemporal rotation symmetry constraints throughout the reconstruction process, enabling more physically accurate modeling of cardiac motion dynamics in cine MRI. In addition, a high-fidelity group filter parameterization mechanism is developed to maintain representation precision while enforcing symmetry constraints. 
Comprehensive experiments on Cardiac CINE MRI datasets demonstrate that DUN-SRE achieves state-of-the-art performance, particularly in preserving rotation-symmetric structures, offering strong generalization capability to a broad range of dynamic MRI reconstruction tasks. 

\end{abstract}


\section{Introduction}
Dynamic Magnetic Resonance Imaging (MRI) is a noninvasive, radiation-free imaging technique that concurrently captures spatial anatomical structures and temporal physiological dynamics~\cite{keall2022integrated}. Among its clinical applications, dynamic cardiac cine MRI with high spatiotemporal resolution plays an important role in the accurate assessment of cardiac morphology and function~\cite{wang2024cine}. However, simultaneously achieving both high spatial and temporal resolution remains a major challenge due to the physical limitations of MRI acquisition. Accelerating MR data acquisition through k-space undersampling has emerged as a promising strategy to mitigate this limitation~\cite{Lustig2007SparseMR}. Yet, reconstructing high-quality images from highly undersampled k-space data poses an inherently ill-posed inverse problem, which necessitates the incorporation of prior knowledge into the reconstruction process~\cite{fessler2010model,liu2020rare}.

Model-driven reconstruction has historically incorporated prior domain knowledge via carefully designed regularization functionals that enforce desired image properties~\cite{fessler2020optimization}. But their performance heavily depends on handcrafted regularization terms that encode analytical priors like sparsity, low-rankness, or smoothness~\cite{zhao2012sparsity,haldar2013rank,He2016lowrk}. 
Such manual approaches often fail to capture the complete statistical structure of real clinical imaging data. With the development of deep learning (DL), data-driven methods have attracted significant attention for their ability to learn implicit image priors directly from data~\cite{hinton2006reducing,lecun2015deep,yu2018aihealth}. Among them, deep unrolling network (DUN) stands out by effectively combining the advantages of both model-driven and data-driven methodologies~\cite{Liang2020deep,hosseini2020RCNMR,cha2020geometric,zhang2022unfoldmr,chen2022aimr}. Specifically, DUNs unroll traditional iterative optimization algorithms into deep neural networks, in which architecture facilitates the seamless integration of helpful domain knowledge like the physical mechanism of MR imaging into the reconstruction process, thereby improving robustness and interpretability. As the core component of DUNs, the regularization term encoding imaging priors is typically implemented through CNN-based parameterization, with network weights optimized end-to-end from training datasets~\cite{qin2018CRNN,kustner2020CINENet,aggarwal2020jmodl,souza2020EnMRprior,martinini2022recovermr,xie2022puert}. By leveraging data-driven, nonlinear imaging priors, DUNs surpass conventional compressed sensing (CS) models in both acceleration factors and reconstruction fidelity. 
A key factor contributing to the effectiveness of CNN-based regularizers is their inherent translational equivariance, \ie,  shifting an input image of CNN is equivalent to shifting all of its intermediate feature maps and output image~\cite{Szegedy2015CNN,Cohen2016GCNN}. This property enables CNN to intrinsically model the translation symmetry, a structural prior universally present in natural and medical images~\cite{Graham2020Dense,lafarge2021medical}. Compared with conventional fully-connected neural networks, this translational equivariance property brings in weight sharing for CNN, which improves parameter efficiency and thus leads to substantially better generalization capability. 

\begin{figure}[!t]
\centering
\includegraphics[width=0.98\linewidth]{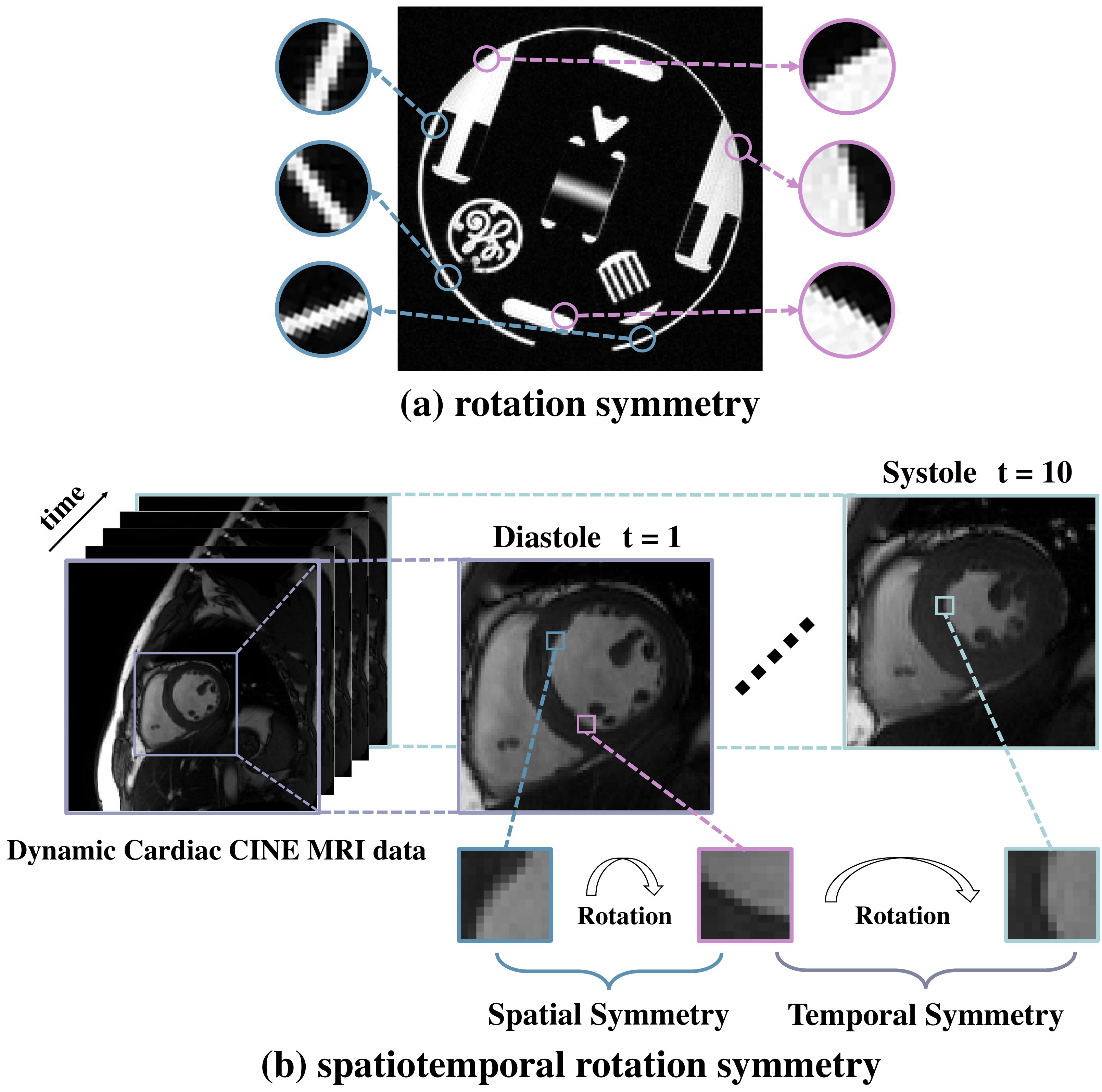}
\caption{(a) An MR phantom image with rotation symmetry pairs indicated by colored circles. (b) An example of dynamic cardiac cine MRI. Similar anatomical structures under different orientations appear within a single frame (spatial symmetry) and across different time frames (temporal symmetry).}
\label{Fig:illustrative_example}
\end{figure}

In addition to translation symmetry, rotation symmetry is another typical structural prior widely observed in general imaging scenarios~\cite{weiler2019general,Wang2021CT,Xie2022Fourier}. As intuitively demonstrated in Fig.~\ref{Fig:illustrative_example}(a), different image patches in an MR phantom image can exhibit similar patterns, though they are in different orientations. To leverage such rotation symmetry priors, a series of Equivariant CNNs (ECNN)~\cite{Weiler2018Steerable,kalogeropoulos2024scale,gerken2023geometric,Chen2023unsupervised} have been developed, which extend regular CNNs to support equivariance under more general transformation groups like rotations. By simply replacing the conventional CNN with ECNN in DUNs, rotational symmetry priors can be readily embedded into the reconstruction process, 
thereby enhancing reconstruction robustness and accuracy. 

While ECNNs provide a theoretically grounded framework for encoding rotation symmetry, current ECNNs are predominantly developed for static imaging scenarios, and their modeling capabilities are largely confined to spatial or volumetric symmetry. However, dynamic MRI data inherently exhibit spatiotemporal symmetry, where similar anatomical structures may appear repeatedly across both spatial locations and time frames with varying orientations. As shown in Fig.~\ref{Fig:illustrative_example}(b), similar myocardial features occur across different frames due to periodic cardiac motion. Effectively capturing such spatiotemporal regularities is crucial for improving dynamic image reconstruction quality. Nonetheless, standard 2D ECNNs are limited to modeling spatial equivariance and fail to account for temporal coherence, while 3D ECNNs are optimized for volumetric data and are suboptimal for time-series medical imaging. These limitations motivate the development of a unified reconstruction framework capable of simultaneously encoding rotation symmetries in both spatial and temporal dimensions. 

To achieve this, we propose a novel Deep Unrolling Network with Spatiotemporal Rotation Equivariance (DUN-SRE) for dynamic MRI reconstruction. The proposed method introduces rotational equivariance in both spatial and temporal dimensions, and preserves this symmetry throughout the unrolled reconstruction process. To the best of our knowledge, this is the first work to develop an ECNN-based model specifically designed for dynamic imaging applications. This work is an extension of our preliminary study~\cite{zhu2024sre}, in which a spatiotemporal ECNN was incorporated as a proximal network within a deep unrolling framework. In this paper, we extend this concept into a more comprehensive framework and make the following contributions: 

\begin{enumerate}
\item 
We design a novel (2+1)D ECNN architecture, termed Spatiotemporal Rotation Equivariant Convolution (SREC), which simultaneously enforces spatial and temporal rotation equivariance for dynamic MRI, enabling explicit modeling of rotation symmetry along both dimensions. Furthermore, the SREC incorporates high-precision group-equivariant filters to maintain anatomical structure fidelity during reconstruction.

\item We develop an end-to-end deep unrolling framework where both the data consistency and proximal mapping modules are implemented using our SREC modules. This unified design ensures strict preservation of rotation equivariance throughout the iterative optimization process, leading to more accurate modeling of non-Gaussian noise distributions and better exploitation of spatiotemporal symmetry priors. 

\item Finally, we evaluate DUN-SRE on both in-house cardiac cine dataset and the publicly available OCMR dataset~\cite{chen2020ocmr} to assess its reconstruction accuracy and generalization capability. It consistently outperforms state-of-the-art dynamic MRI reconstruction methods in terms of both qualitative visual quality and quantitative metrics, particularly at high acceleration factors. 
\end{enumerate}

\section{Related Work}
\subsection{Equivariant CNN}
Given that physiological organ motion in Euclidean space exhibits inherent transformation symmetries, maintaining network equivariance to these symmetry groups is fundamental for physically plausible dynamic MRI reconstruction. 
Recently, a series of equivariant CNNs~\cite{weiler2019general,shen2020pdo,cesa2022En} have been developed to incorporate equivariance directly into the network architecture, aiming to model various transformation symmetry priors inherently existing in images, \eg, reflection, scaling and rotation symmetries. 
In geometric deep learning, equivariance refers to the fundamental property where transforming an input (\eg, rotating an image) generates an equivalently transformed output (\eg, rotating the features), preserving structural relationships~\cite{weiler2019general,cesa2022En}. Specifically, let $\Psi$ be a mapping from the input feature space to the output feature space, and $\mathcal{G}$ is a group transformations. $\Psi$ is equivariant with respect to the action of $\mathcal{G}$, if for any $g\in\mathcal{G}$,
\begin{equation}
\Psi(\rho_{g}(\bm{x})) = \rho'_{g}(\Psi(\bm{x})), 
\label{Related Work Equivariance Definition}
\end{equation}
where $\bm{x}$ denotes an input feature map, and $\rho_{g}(\cdot)$ and $\rho'_{g}(\cdot)$ represent how the transformation $g$ acts on input and output features, respectively.

Early attempts to exploit transformation symmetry prior in images mainly focus on rotations in 2D Euclidean space, \eg, GCNN~\cite{Cohen2016GCNN} and HexaConv~\cite{hoogeboom2018hexaconv} construct equivariant convolutions for $\pi/2$ and $\pi/3$ rotations, respectively. In~\cite{worrall2017harmonic}, to leverage more rotational symmetries, feature maps and filters at different orientations are generated using bilinear interpolation, resulting in inherently approximate equivariance. Equivariant networks have been applied not only to exploit planar symmetry priors, but also extended to 3D voxel grids for modeling volumetric symmetries. CubeNet~\cite{worrall2018cubenet} introduces a cube-like symmetry structure that enables linear equivariance to 3D rotations and translations on voxelized data, thereby preserving both global and local shape representations of 3D objects. In addition, 
Cohen \etal~\cite{cohen2018spherical} developed spherical CNNs to achieve exact $\mathrm{SO}(3)$-equivariance for data defined on spherical manifolds, effectively addressing the distortion artifacts inherent in planar projections of 3D spherical data. 

While existing equivariant CNNs successfully exploit planar and volumetric rotation symmetries, they fundamentally lack mechanisms for temporal rotation equivariance in dynamic imaging (2D+t). This represents a critical limitation for dynamic MRI reconstruction, where spatiotemporal rotation symmetries naturally exist in cardiac motion patterns and could substantially enhance reconstruction fidelity if properly modeled. Hence, an ideal framework should incorporate continuous temporal rotation equivariance with spatial symmetries.

\subsection{Deep Unrolling Network for MRI Reconstruction}
The dynamic MRI acquisition follows the forward model, 
\begin{equation}
\bm{y} = \bm{A}\bm{x} + \delta,
\label{Eq rw1}
\end{equation}
where $\bm{x}$ represents the dynamic MRI sequence to be reconstructed, $\bm{y}$ denotes the observed $k$-space data; $\delta$ represents the measurement error which can be well modeled as noise; and $\bm{A}: \mathbb{C}^{N} \to \mathbb{C}^{M}$ is an undersampled Fourier encoding matrix, with $M \leq N$. 
Commonly, $R \triangleq N \text{/} M$ is referred to as the acceleration rate. 
The image reconstruction can be formulated as a regularized optimization problem:
\begin{equation}
\hat{\bm{x}}=\argmin_{\bm{x}} \left[ \mathcal{M}\left(\bm{Ax}, \bm{y}\right) + \lambda R(\bm{x}) \right], 
\label{Eq rw2}
\end{equation}
where $\lambda$ is the trade-off parameter; the data consistency (DC) term $\mathcal{M}(\bm{Ax}, \bm{y})$ guarantees data fidelity, and the regularization term $R(\bm{x})$ depends on the prior knowledge of the input $\bm{x}$. 

In recent years, DUNs have shown great potential in solving the problem in Eq.~\eqref{Eq rw2} for MRI reconstruction, by combining the interpretability of traditional optimization with the learning capacity of deep neural networks. To leverage domain-specific knowledge, such as Fourier transform and spatiotemporal correlations in dynamic MR imaging, DUNs integrate the iterative steps of an optimization algorithm into neural networks, \eg, alternating direction method of multipliers network~\cite{yang2018admm}, iterative shrinkage-thresholding algorithm network~\cite{zhang2018ista}, and variational network~\cite{hammernik2018VN}. 

To further improve reconstruction fidelity, many methods~\cite{zheng2019deepDC,gungor2022transms,liu2023DCprior,gungor2023deq} have been recently developed to learn an implicit DC term using CNNs, thereby adapting a more precise probability distribution of system noise. Traditional approaches typically define the DC term explicitly as an $L_2$-norm of the difference between the reconstructed and observed $k$-space data, \ie, $\| \bm{Ax}-\bm{y}\|_{2}^{2}$, and assume the noise $\delta$ follows an uncorrelated additive normal distribution~\cite{brown2014magnetic,ahmad2020plug}. However, this idealized noise model proves inadequate in practice, as real MRI systems exhibit more complex noise characteristics that deviate significantly from simple additive Gaussian assumptions. In contrast, DUNs adopt an alternative assumption in which the measurement noise still satisfies an exponential distribution after a nonlinear transformation, thereby enabling better adaptability to the complex noise characteristics encountered in real-world MRI systems~\cite{cheng2021LDC}. 

Recently, Celledoni \etal~\cite{celledoni2021inverse} revealed a fundamental design principle for DUNs: the explicit incorporation of physically meaningful transformation symmetries inherent to the MR reconstruction problem as inductive biases in network architecture. Since then, a series of methods have purposefully employed equivariant CNNs in deep unrolling networks to incorporate transformation symmetry priors. For example, Gunel \etal~\cite{gunel2022scale} implemented the unrolling network using scale-equivariant convolutional layers, thereby enhancing robustness against anatomical size variations and field-of-view inconsistencies commonly encountered in multi-scanner clinical MRI datasets. 
To explore alternative transformations, Fu \etal~\cite{fu2024proximal} embed planar rotation symmetry priors into the DUNs via a rotation equivariant proximal network, leading to a more accurate representation of the image prior and improved model generalization ability. 


A fundamental limitation persists in current deep equivariant unrolling networks: their architectures predominantly adopt existing equivariant modules through heuristic integration, without rigorous analysis of how intrinsic image priors propagate theoretically through the unrolling process. Our work addresses this gap by systematically investigating the incorporation of temporal rotation symmetry, one of the most universal yet underutilized priors in dynamic MRI, into the unrolling framework through both theoretically sound and computationally tractable means. 

\section{Methodology}

In this section, we present DUN-SRE, our deep unrolling framework with spatiotemporal rotation equivariance. We first formulate the DUN for MRI reconstruction, then introduce the SREC architecture that effectively exploits rotation symmetry in both spatial and temporal dimensions. Finally, we analyze equivariance in the data consistency term and implement fully rotation-equivariant proximal and DC networks using SREC, ensuring end-to-end equivariance across all unrolled iterations.

\subsection{Problem formulation}
We leverage recent advances in deep unrolling methods and adopt the Proximal gradient descent (PGD) algorithm~\cite{parikh2014proximal}, in which both the DC term (data likelihood) and regularization term (prior) are learned through deep neural networks. The optimization problem in Eq. \eqref{Eq rw2} is iteratively solved by alternating between a DC update and a proximal update, formulated as:
\begin{equation}
\begin{cases}
\bm{x}^{k+1/2} = \bm{x}^{k} - \eta \nabla_{\bm{x}}\mathcal{M}(\bm{Ax}, \bm{y}) , \\
\bm{x}^{k+1} = \mathcal{H}^{k} (\bm{x}^{k+1/2},\theta^{k})
\end{cases}
\label{Eq rw3}
\end{equation}
where $\bm{x}^{k+1/2}$ denotes the intermediate result after the DC update, and $\bm{x}^k$ is the reconstruction at the $k^\text{th}$ iteration; $\eta$ is the step size; $\nabla_{\bm{x}}$ represents the gradient of $\mathcal{M}\left(\bm{Ax}, \bm{y}\right)$ with respect to $\bm{x}$; and $\mathcal{H}^{k}$ is a proximal network with iteration-specific parameters $\theta^{k}$, which is a parameterized proximal operator associated with the regularization term $R(\cdot)$. 

It is easy to find that the image prior knowledge is indeed primarily reflected in proximal network during optimization and is learned implicitly from training data. At the same time, the DC term is also learned in a data-driven manner, enabling a more accurate modeling of data fidelity under realistic noise distributions.

\subsection{(2+1)D Spatiotemporal Rotation Equivariant Convolutions}
Most of the existing DU methods for MR dynamic reconstruction tend to leverage the dynamic information and spatiotemporal correlations inherently existing in the data to facilitate reconstruction. In dynamic MR sequences, a large number of similar patterns are present under varying conditions across different temporal frames, reflecting abundant temporal symmetry priors associated with various transformations. This motivates us to incorporate these useful image priors into DU framework, thereby further enhancing improving reconstruction performance. However, current ECNNs are limited in capturing the temporal symmetry priors of certain transformations. 

\begin{figure}[!t]
\centering
\includegraphics[width=0.98\linewidth]{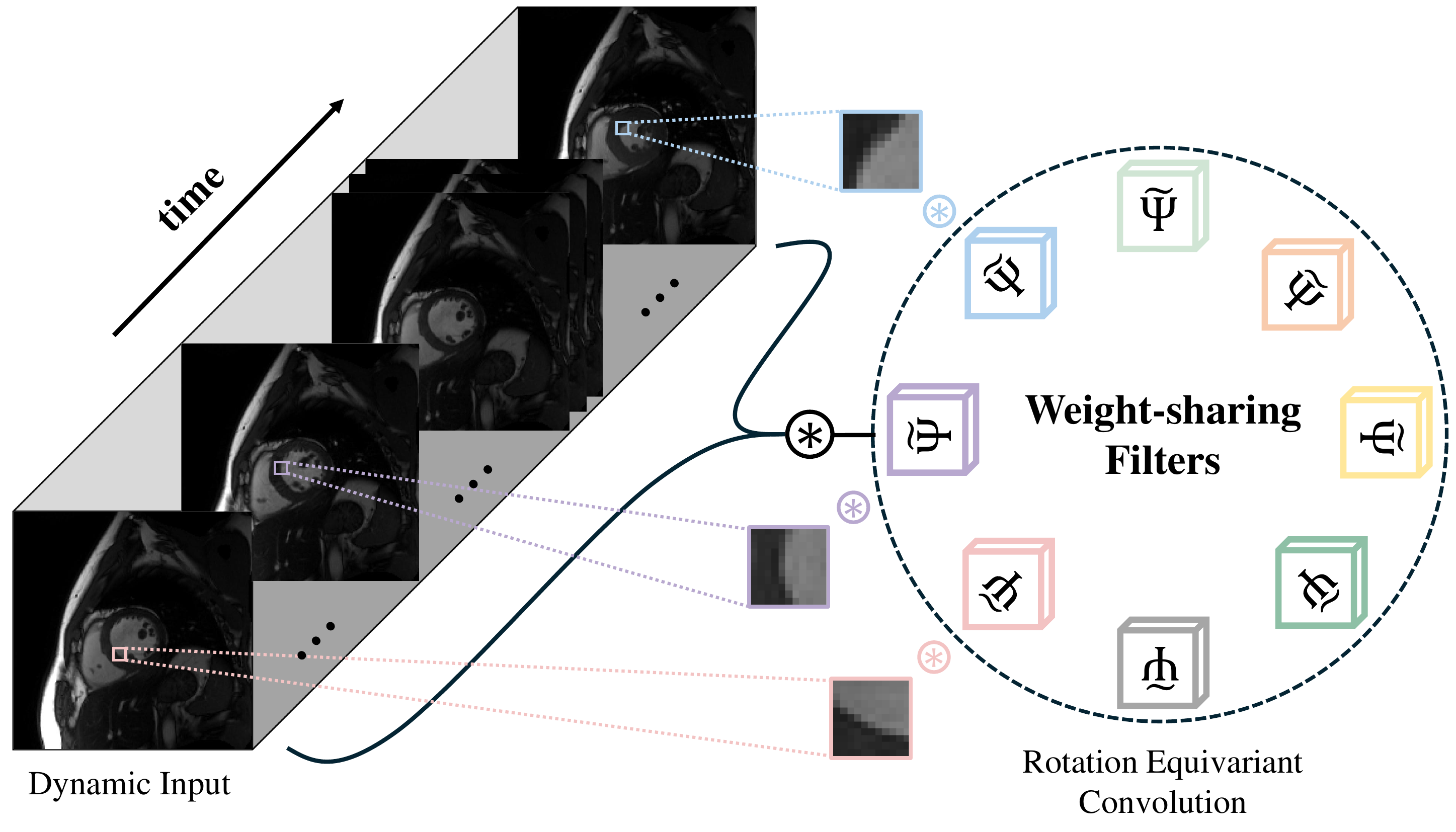}
\caption{A group of weight-sharing SREC filters can extract all similar patterns across different frames in dynamic MR data, where each rotated kernel corresponds to features at a specific orientation.}
\label{Method_2_TS}
\end{figure}

To address this, we propose a SREC module, which aims to robustly establish rotation equivariance within dynamic reconstruction models. Specifically, SREC constructs a set of weight-sharing convolutional filters that can be reused to extract similar patterns under different orientations across all frames of the dynamic sequence, as depicted in Fig. \ref{Method_2_TS}. Benefiting from this design, SREC can effectively exploit rotation symmetry priors along the temporal dimension, with only a single filter weight needing to be learned. The proposed SREC consists of four basic building blocks: the input, intermediate, output and temporal equivariant layer, which will be introduced in the following sections.

\begin{table}[!t]
  \caption{The involved concepts and notations for equivariant convolutions in the continuous and discrete domains.\\}\label{Notations}
  \centering
  \footnotesize
      \begin{tabular}{c c c}
        \hline 
        \hline 
        \vspace{-7pt}\\ 
         Concept & \hspace{10em} Notation & \\ 
                & Continuous & Discrete \\
        \hline  
        \vspace{-7pt} \\
        Input Image & $f(\bm{r},t)$ & $I$ \\
        Transformation Group & $\mathrm{SO}(2)$ & $G$ \\
        Group element/Index & $A, B \in \mathrm{SO}(2)$ & $A, B \in G $\\
        Trivial Group element & $D \in \mathrm{SO}(2)$ & $D \in \mathit{SG}$\\
        \vspace{2pt} Feature Map & $e(\bm{r},A,t)$ & $F^A$ \\ 
        \hline  
        \vspace{-7pt} \\
        Filter (Input) &  $\varphi_{in}(A^{-1}\bm{r},t)$  &$\tilde{\Psi}^{A}$  \\
        Filter (Intermediate) &  $\varphi_{A}(B^{-1}\bm{r},t)$   &$\tilde{\Phi}^{B,B^{-1}A}$ \\
        Filter (Temporal) & $\varphi_{B}(\bm{r},t,D^{-1})$ &  $\tilde{\Lambda}^{D,D^{-1}B}$  \\
        \vspace{2pt} Filter (Output) &   $\varphi_{out}(D^{-1}\bm{r},t)$ &$\tilde{\Upsilon}^{D}$ \\
        \hline \hline
    \end{tabular}
\end{table}

\subsubsection{Necessary Notations and Concepts}
For clarity, the major notations correspondingly used in continuous and discrete domains are listed in Table \ref{Notations}. In the continuous domain, we represent the input dynamic image sequence as a function $f:\mathbb{R}^2 \times \mathbb{R} \to \mathbb{R}$, denoted as $f(\bm{r},t)$, where $\bm{r}=(x,y)$ corresponds to the spatial domain, and $t$ to the temporal dimension. We consider the planar equivariance of $\bm{r}$ on a special Euclidean group $\mathrm{SE}(2)$, which is the group of isometries of the plane $\mathbb{R}^2$. Formally, $\mathrm{SE}(2)$ can be viewed as the semidirect product of the translation group $(\mathbb{R}^2, +)$ and the special orthogonal group $\mathrm{SO}(2)$ that represents a rotation group without reflection, expressed as $\mathrm{SE}(2) \cong (\mathbb{R}^2, +) \rtimes \mathrm{SO}(2)$. We use $A$ to parameterize $\mathrm{SO}(2)$, and represent an element of $\mathrm{SE}(2)$ as $(\bm{r},A,t)$. By restricting the domain of $\bm{r}$ and $A$, we can also use this representation to parameterize any subgroup of $\mathrm{SE}(2)$. The intermediate feature map can be model as a function defined on $\mathrm{SE}(2)$, denoted as $e(\bm{r},A,t)$.

In practice, convolution operations are generally implemented in the discrete domain. The input digital image $I \in \mathbb{R}^{x\times y\times t}$ can be viewed as a three-dimensional discretization of the smooth function $f(\bm{r},t)$ at the cell-center of a regular grid. Note that in rotation equivariant networks, a feature map is a tensor (i.e., multi-channel matrix), with $F \in \mathbb{R}^{x\times y\times t \times N_G}$, where the fourth mode is with respect to the rotation group $G$, and $N_G$ denotes the number of elements in $G$. The group $G$ represents a subgroup of $\mathrm{SO}(2)$, and it is also regarded as the discretization of $\mathrm{SO}(2)$ in this paper. Specifically, $G$ describes discrete in-plane rotations with evenly spaced orientations $G=\left\{ A_{k,\theta} \mid \theta = \frac{2\pi k}{N_G},\; k = 1, 2, \cdots, N_G \right\}, k \in \mathbb{N}_{+}$, where $k$ indexes the corresponding rotation operation in the group. A single-channel $p\times p \times p$ 3D convolutional filter can be obtained by discretizing a continuous filter $\varphi:\mathbb{R}^2 \times \mathbb{R} \to \mathbb{R}$. Accordingly, we define the dicrete filters for input, intermediate, temporal and output layers as $\tilde{\Psi}\in \mathbb{R}^{t\times p\times p\times N_G}$, $\tilde{\Phi}\in \mathbb{R}^{t\times p\times p\times N_G\times N_G}$, $\tilde{\Lambda}\in \mathbb{R}^{t\times p\times p\times N_G\times N_G}$, and $\tilde{\Upsilon}\in \mathbb{R}^{t\times p\times p\times N_G}$, respectively.
\begin{figure*}[!t]
\centering
\includegraphics[width=\textwidth]{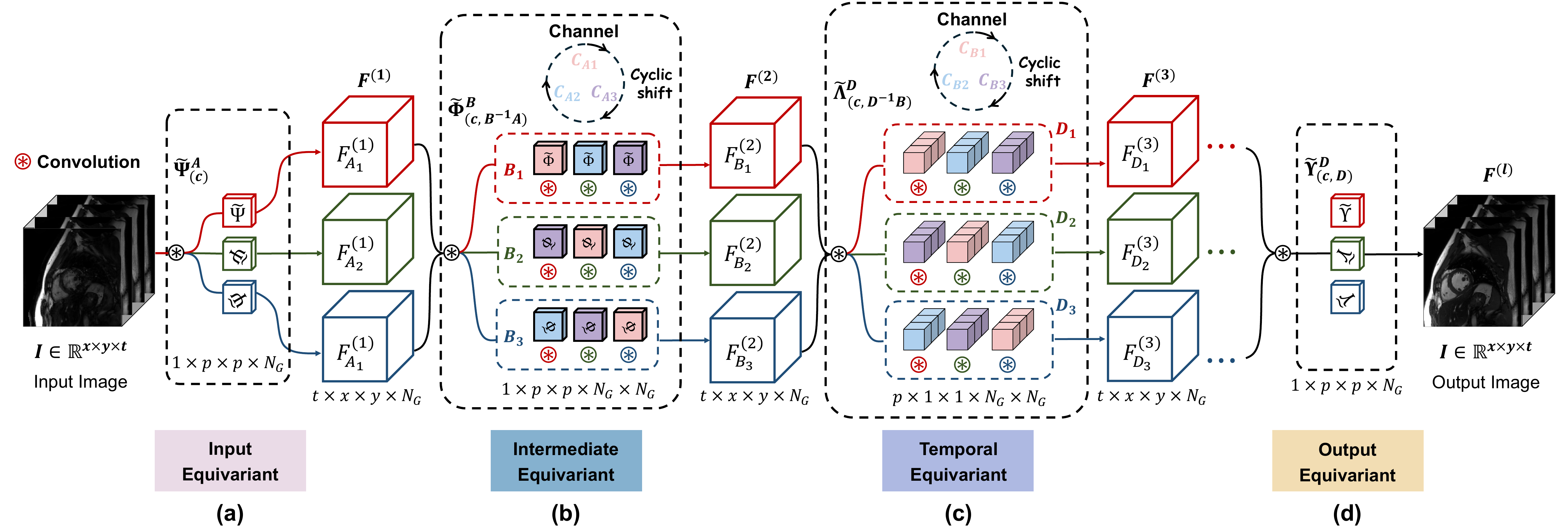}
\caption{Basic structure of an example network constructed using proposed SREC for dynamic MR reconstruction, where we set the transformation group $G (A_{k},B_{k}\in G)$ and $\mathit{SG}(D_{k}\in\mathit{SG})$ as $2\pi k/3$ rotations, $k=1,2,3.$. For clarity, only a single feature channel along with its corresponding group element is shown. (a)-(d) represents the spatiotemporal equivariant convolutions of the input layer, intermediate layers, temporal layer and output layer, respectively. The superscript $(l)$ of $F$ indicates the layer index of the feature map.}
\label{Method_1_SREC}
\end{figure*}

\subsubsection{Input Equivariant Layer}
The equivariant CNNs are designed based on the observation that when the relative orientation of an image and a convolution filter changes, the extracted features will be different. To ensure the feature robustness to input image rotation, the input equivariant layer creates a set of weight-sharing convolution filters by copying and rotating an initial base filter on the spatial plane of $\bm{r}$, where the rotations $A$ are the elements of the group $G$. These filters are then sequentially convolved with the input image to produce $N_{G}$ feature maps, each corresponding to a specific orientation of the target feature. Consequently, the extended channels of the resulting feature map $F \in \mathbb{R}^{x\times y\times t \times N_{G}}$ are arranged successively in alignment with the group elements, which can be viewed as a group of sub-channels. This process is illustrated in Fig. \ref{Method_1_SREC}(a), where rotated Greek letters indicate the spatial orientations of the filters, and each color corresponds to a specific rotation angle in $G$. The $A_{k} \in G$ is a rotation operation, and is also used to represent the index of the sub-channel within each group, expressed as $F_{A_{k}}$. 

In the continuous domain, the convolution filter for input equivariant layer is denoted by $\Psi$, which maps an input function $f\in \mathcal{C}^\infty(\mathbb{R}^2 \times \mathbb{R})$ to a set of feature maps defined on $\mathrm{SE(2)} \rtimes t$. Specifically, for any $(\bm{u}, A, t)\in \mathrm{SE(2)} \rtimes t$, 
\begin{equation}
\Psi\left[f\right](\bm{u}, A, t) = \int_{\mathbb{R}^2} \varphi_{in}(A^{-1} \bm{r},t) \, f(\bm{u} - \bm{r},t) \, d\sigma(\bm{r}),
\label{Eq input_continuous}
\end{equation}
where $\sigma$ is a measure on $\mathbb{R}^2$, and $\varphi_{in}$ represents the continuous filter of the input layer. Consequently, the resulting feature map $F$ can be represented as a smooth function $e:\mathbb{R}^{2}\times \mathbb{R} \times G \to \mathbb{R}$.

For digital images, we consider an input with $C$ channels, denoted as $I \in \mathbb{R}^{x\times y\times t \times C}$, where $c=1,\ldots,C$. For any $A\in G$, the convolution of $I$ and $\tilde{\Psi}^{A}$ can be calculated by discretizing Eq. (\ref{Eq input_continuous}), which is:
\begin{equation}
({\tilde{\Psi}\star I})^{A} = \sum\limits_{c} \tilde{\Psi}_{(c)}^{A}\ast I_{c}.
\label{Eq input_discrete}
\end{equation}
where $\star$ represents the group convolution operator, $\ast$ is the standard discrete convolution. 

\subsubsection{Intermediate Equivariant Layer}
Since the input feature map to the intermediate layer is now a function on $G$, we must ensure that the filters in the intermediate layer also give a function on $G$. Similar to the input layer, we perform planar rotations $B \in G$ on each base filter to generate $N_{G}$ rotated copies, which are then sequentially convolved with the input feature map $F$. Notably, when rotating the filters, an additional position shift on channel must be cyclically performed, in line with the associated group action. An intuitive illustration of this process is provided in Fig. \ref{Method_1_SREC}(b). In the intermediate layer, the convolution filters expanded by the rotation group $B$ are represented by the colored box with dashed lines. Each filter contains three sub-channels $C_{A_k}$, where each sub-channel corresponds to a feature map channel extended by the rotation group from the previous layer. In this case, the previous layer is the input equivariant layer, and thus the corresponding rotation group is $A$. The kernels of sub-channels with the same color (pink, light blue, and purple) share the same weight. The color of each convolution operator shown below the sub-channels indicates which feature map channel from the previous layer it should be convolved with.

In the continuous domain, the convolution filter for intermediate layer is denoted as $\Phi$, which maps a feature map $e\in \mathcal{C}^\infty(\mathrm{SE(2)} \rtimes t)$ to another feature map defined on $\mathrm{SE(2)}\rtimes t$. Specifically, for any $(\bm{u}, B, t)\in \mathrm{SE(2)}\rtimes t$,
\begin{equation}
\begin{split}
&\Phi\left[e\right](\bm{u}, B, t)\\
&=\int_{\mathrm{SO}(2)} \int_{\mathbb{R}^2} \varphi_{A}(B^{-1}\bm{r},t) \, e(\bm{u} - \bm{r},BA,t) \, d\sigma(\bm{r} )dv(A),
\end{split}
\label{Eq intermediate_continuous}
\end{equation}
where $v$ is the Haar measure on $\mathrm{SO}(2)$, $A,B \in \mathrm{SO}(2)$ indicate orthogonal transformations in the considered group, and $\varphi_{A}$ denotes the filter with respect to the channel of feature map indexed by $A$. 

In the discrete domain, the convolution of $F$ and $\tilde{\Phi}$ can be calculated by discretizing Eq. (\ref{Eq intermediate_continuous}), expressed as
\begin{equation}
({\tilde{\Phi}\star F})^{B}= \sum\limits_{c,\, A} {\tilde{\Phi}_{(c,\, B^{-1}A)}^{B}}\ast F_{(c,\, A)}^{B}.
\label{Eq intermediate_discrete}
\end{equation}
where $c$ denotes the channel index, and $(c,B^{-1}A)$ is the index of sub-channel within the $c^{th}$ channel. As previously described, each of the $c$ channels has been extended by the group $A$, and thus contains $N_{G}$ sub-channels, resulting in a total of $c\times N_{G}$ channels.

\subsubsection{Temporal Equivariant Layer}
Since equivariant convolutions rely on the repeated exploitation of similar or identical patterns, those pairs of rotation-symmetric patterns are expected to remain stable within the 2D plane of each frame. Therefore, we adopt a (2+1)D convolution structure~\cite{tran2018Conv21D} to construct our SREC, which factorizes a 3D convolution $d\times d\times d$ into two separate and successive operations, a 2D spatial convolution $1\times d\times d$ and a 1D temporal convolution $d\times 1\times 1$. The input and intermediate equivariant layers are constructed using the 2D spatial convolutions, which facilitate planar stability when extracting the rotation-symmetric pattern pairs. However, the symmetry priors embedded in these patter pairs still cannot be exploited by the network, as no effective mechanism is established to transfer equivariance between the 2D convolutional layers. 

To address this problem, we specifically design a novel 1D temporal equivariant convolutional layer to enable effective equivariance transfer between any two 2D convolutional layers, thereby preserving the overall equivariance of the network. Specifically, we define a special group $\mathit{SG}$ by restricting the temporal layer function to the cosets of the group $G$, and construct the temporal equivariant layer based on the group $\mathit{SG}$. Although $\mathit{SG}$ is theoretically defined as a rotation group, it acts as a trivial group in practical implementation, which is composed of only a single element, i.e., the identity.This is because the temporal equivariant layer practically operates as a pseudo-1D convolution with a spatial dimension of $1\times 1$, rendering the explicit planar rotation of $\bm{r}$ unnecessary. Nevertheless, to maintain equivariance between two 2D layers, it remains essential to perform a cyclic shift of sub-channels within each filter. We use $D$ to parameterize the group $\mathit{SG}$. The architecture of the temporal layer is illustrated in Fig. ~\ref{Method_1_SREC}(c). 
 
In the continuous domain, the convolution for temporal layer is denoted as $\Lambda$, which maps a feature map $e\in \mathcal{C}^\infty(\mathrm{SE(2)} \rtimes t)$ to another feature map defined on $\mathrm{SE(2)}\rtimes t$. Specifically, for any $(\bm{u}, D,t)\in \mathit{SG}$, the operation is given by:
\begin{equation}
\begin{split}
&\Lambda\left[e\right](\bm{u}, D, t) \\
&= \int_{\mathrm{SO}(2)} \int_{\mathbb{R}} \varphi_{B}(\bm{r}, t,D^{-1}) \, e(\bm{r},DB,t-\tau) \, d\sigma(t)dv(B),
\end{split}
\label{Eq temporal_continuous}
\end{equation}
$\varphi_{B}$ denotes the filter with respect to the channel of feature map indexed by $B$. 

In the discrete domain, the convolution of $F$ and $\tilde{\Lambda}$ can be calculated by discretizing Eq. (\ref{Eq temporal_continuous}), expressed as:
\begin{equation}
({\tilde{\Lambda}\circledast F})^{D}= \sum\limits_{c,B} {\tilde{\Lambda}_{(c,\, D^{-1}B)}^{D}} \ast F_{(c,\, B)}^{D} .
\label{Eq temporal_discrete_withchannels}
\end{equation}
where $\circledast$ denotes the special group convolution defined on $\mathit{SG}$, and $(c,D^{-1}B)$ indexes the sub-channel within the $c^{th}$ channel. The inverse operation $D^{-1}$ is applied to cyclically shift the channels indexed by the rotation group $B$ from the previous layer.

The (2+1)D convolutional structure first achieves spatial equivariance through 2D input and imtermediate layers, and then establishes temporal equivariance using the proposed 1D temporal equivariant layer in a decomposed manner. Through the temporal equivariant layer, convolutional filters within the same group can be shared across all frames to extract similar patterns under different orientations, thereby fully exploiting rotation symmetry priors along the temporal dimension, as intuitively depicted in Fig. \ref{Method_2_TS}. Consequently, the proposed temporal layer extends the scope of available symmetry priors from a single frame to the entire dynamic sequence, thereby significantly improving the parameter efficiency of the network. 
\begin{figure*}[!t]
\centering
\includegraphics[width=\textwidth]{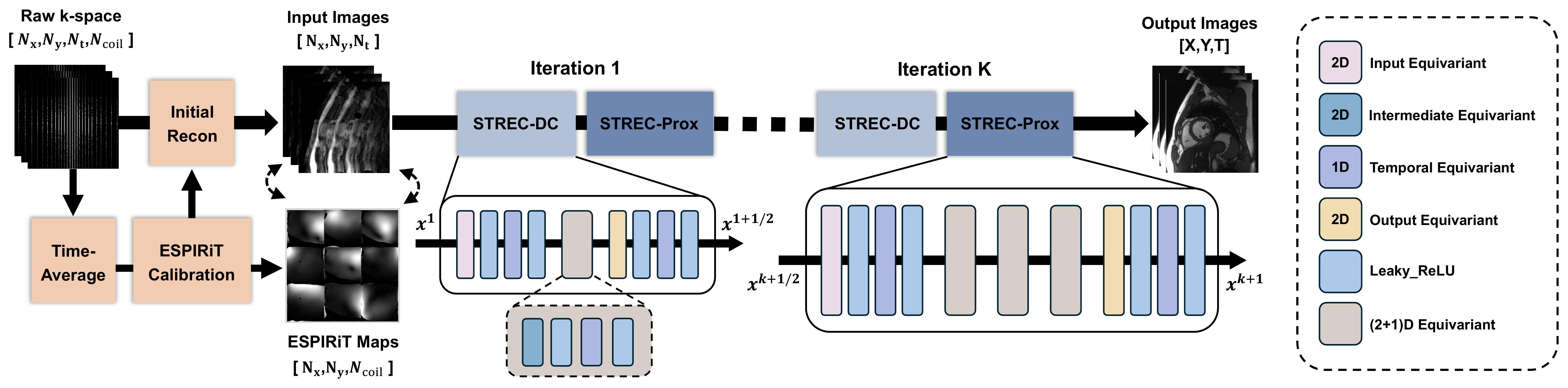}
\caption{The full pipeline of DUN-SRE for dynamic MR reconstruction. A fully sampled calibration region is extracted from the time-averaged k-space data and used to compute the ESPIRiT sensitivity maps. These maps are used to compute an initial zero-filled reconstruction from the raw k-space data with complex-valued channels. The network is optimized iteratively by alternating between the SREC-DC network and the SREC-Proximal network.}
\label{Method_3_DUN}
\end{figure*}

\subsubsection{Output Equivariant Layer}
In contrast to the output equivariant layers used in high-level tasks, which typically apply group pooling over the orientation dimension, a specialized output layer is adopted for dynamic MR reconstruction in our work. To guarantee equivariance while avoiding the loss of fine texture details in images, our output layer reduces the number of channels expanded by the rotation group used in the previous layer and simultaneously maintains consistency between the output and input image channels. Specifically, we use $\Upsilon$ to denote the convolution on the final layer, which maps a feature map $e\in \mathcal{C}^\infty(\mathrm{SE(2)} \rtimes t)$ to a function in $\mathcal{C}^\infty(\mathbb{R}^2 \times \mathbb{R})$. For any $(\bm{u},t)\in \mathbb{R}^2\times \mathbb{R} $, we define:
\begin{equation}
\begin{split}
&\Upsilon\left[e\right](\bm{u}, t) \\
&= \int_{\mathrm{SO}(2)} \int_{\mathbb{R}^2} \varphi_{out}(D^{-1}\bm{r},t) \, e(\bm{u} - \bm{r},D,t) \, d\sigma(\bm{r} )dv(D),
\end{split}
\label{Eq output_continuous}
\end{equation}
where $D\in \mathit{SG}$. The discretization of Eq.(\ref{Eq output_continuous}) can be expressed as
\begin{equation}
({\tilde{\Upsilon}\star F})= \sum_{c,\, D \in \mathit{SG}} \tilde{\Upsilon}_{(c,D)}^{D}\ast F_{(c,D)}^{D}.
\label{Eq output_discrete}
\end{equation}
One can view the output equivariant layer in Fig. \ref{Method_1_SREC}(d) to easily understand this progress.

\subsection{Spatiotemporal Rotation Equivariant Unrolling Network}
\subsubsection{Spatiotemporal Equivariant Proximal Network}
We then design a proximal network using SREC to insightly embed spatiotemporal rotation symmetry priors, referred to as SREC-Prox. Since the convolution operators are inherently translation equivariant, it is easy to deduce that the proposed convolutions $L_{(\cdot)}$ are equivariant under joint rotations and translations in the group $\mathrm{SE(2)}$, which can be expressed as,
\begin{equation}
\begin{split}
\Psi \left[ \rho_{\tilde{G}}^{R} \left[ f\right] \right]&= \rho_{\tilde{G}}^{E} \left[ \Psi \left[ f\right] \right],\\
\Phi \left[ \rho_{\tilde{G}}^{E} \left[ e\right] \right]&= \rho_{\tilde{G}}^{E} \left[ \Phi \left[ e\right] \right],\\
\Lambda \left[ \rho_{\tilde{\mathit{SG}}}^{E} \left[ e\right] \right]&= \rho_{\tilde{\mathit{SG}}}^{E} \left[ \Lambda \left[ e\right] \right],\\
\Upsilon \left[ \rho_{\tilde{G}}^{E} \left[ e\right] \right]&= \rho_{\tilde{G}}^{R} \left[ \Upsilon \left[ e\right] \right],
\end{split}
\label{Eq Equivariance of layers}
\end{equation}
where the rotation operators $\rho_{\tilde{G}}^{R}$ and $\rho_{\tilde{G}}^{E}$ are defined on $\mathcal{C}^\infty(\mathbb{R}^2)$ and $\mathcal{C}^\infty(\mathrm{SE(2)})$ respectively. 

Furthermore, it has been proven that both the linear layers and nonlinear activation functions (\eg, LeakyReLU function) are equivariant~\cite{Cohen2016GCNN,Weiler2018Steerable}. Therefore, the combination of each proposed convolutional layer and its activation function also forms an equivariant mapping, so that the SREC-Prox, constructed by stacking these mappings, guarantees the overall equivariance of the entire network. The detailed architecture of SREC-Prox is shown in Fig.~\ref{Method_3_DUN}, which is denonted as $\mathcal{H}_{\operatorname{SREC}}$. 

\subsubsection{Equivariant Data Consistency Network}
In most deep unrolling approaches, the DC update not only plays a crucial role in retaining essential information from raw measurements, but also has a substantial impact on the overall transformation equivariance of the unrolling model. As shown in Eq. (\ref{Eq rw3}), the DC update and the proximal network are performed in a sequentially alternating manner during iterative optimization. Therefore, only when both the DC component and the proximal network possess the equivariance property can the symmetry priors be consistently propagated through the unrolling process. In fact, the traditional DC term in the form of the $L_2$-norm can be theoretically demonstrated to inherently possess both translational and rotational equivariance. Formally, by treating the image as an element in a Hilbert space, we have the following conclusion:
\begin{rmk}\label{remark 1}
Suppose that $\mathcal{X}$ is a Hilbert space and $\rho$ is a unitary transformation of a group $\mathcal{G}$ on $\mathcal{X}$. if the data consistency term follows an explicit $L_2$-norm formulation, then the corresponding data consistency update $\mathcal{Q}_{\operatorname{DC}}$ is equivariant to the unitary transformation $\rho$, in sense that
\begin{equation}\label{Eq remark 1}
\mathcal{Q}_{\operatorname{DC}}[\rho_g](X) = \rho_g[\mathcal{Q}_{\operatorname{DC}}](X), \quad \forall X \in \mathcal{X}, \ g \in \mathcal{G}
\end{equation}
\end{rmk}

According to Remark~\ref{remark 1}, it can be inferred that a key principle in designing a DC item is to ensure its equivariance to translation or rotation. This fully complies with our common sense that when the input image rotates, the target image will also rotate accordingly to enforce the minimal error.

However, all existing deep DC methods are implemented using conventional CNNs, which inherently lack the ability to preserve rotational equivariance. As a result, the conventional CNN-based DC component fails to maintain rotational equivariance between successive unrolled iterations, ultimately destroying the overall equivariance of the unrolling model. To address this problem, we construct an equivariant DC network using our proposed SREC, referred to as SREC-DC, whose detailed architecture is as shown in Fig. \ref{Method_3_DUN}. This equivariant DC network not only learns a more precise and realistic noise distribution but also extends equivariance to the entire unrolled network, rather than restricting it to the proximal component alone. Given that both the SREC-proximal network and the SREC-DC network exhibit rotational equivariance, the completed deep unrolling method, DUN-SRE, achieves global rotational equivariance, formally expressed as:
\begin{equation}
\label{Overall Equivariance}
\resizebox{0.95\hsize}{!}{$
\begin{split}
\mathcal{H}_{\operatorname{SREC}} \left( \mathcal{Q}_{\operatorname{SREC}}\left[\rho_g \right]\left(X^{k}\right) \right) &=
\mathcal{H}_{\operatorname{SREC}} \left( \rho_g \left[\mathcal{Q}_{\operatorname{SREC}} \right]\left(X^{k}\right) \right) \\
&= \rho_g  \left(\mathcal{H}_{\operatorname{SREC}} \left[\mathcal{Q}_{\operatorname{SREC}}\right]\left(X^{k}\right)\right)
\end{split}
$}
\end{equation}
where $\mathcal{Q}_{\operatorname{SREC}}$ denotes the equivariant DC network constructed with SREC.

The pipeline of the DUN-SRE dynamic MR reconstruction model is illustrated in Fig. \ref{Method_3_DUN}. Specifically, DUN-SRE takes as inputs a zero-filled reconstruction of a 2D cardiac cine slice along with its corresponding ESPIRiT maps computed from the time-averaged k-space data. The network is trained to reconstruct images that closely match the corresponding fully sampled ground truth, guided by an $\ell_1$ pixel-wise loss.

\subsection{Filter Parametrization}
Furthermore, we introduce a high-precision filter parametrization strategy to improve the accuracy of convolution filter representation. The basic idea is to define the objective functional filter $\psi$ as the linear combination of a set of basis functions $\left \{ \psi_n \right \} _{n=1}^{N}$, which can be formulated as: 
\begin{equation}
\psi(\bm{s}) =\sum_{n=1}^{N}{w_n\psi_n}(\bm{s}),
\label{Filter Parametrization}
\end{equation}
where $\bm{s}=[s_1,s_2]^T$ represents the two spatial coordinates, $N$  denotes the number of basis functions, $w_n$ the n-th coefficient. Instead of directly learning the filter weights, the network learns the coefficients $w_n$ of these basis functions during training. Given that the Fourier series expansion is equivariant to the inverse discrete Fourier transform and introduces zero representation error, we adopt 1D Fourier series as basis functions to construct filters for the temporal equivariant layer, and use 2D Fourier basis functions for the input, intermediate and output equivariant layers. Specifically, the filters in the 1D temporal equivariant convolutions are constructed as:
\begin{equation}
\tilde{\phi}(\bm{s}) = \sum_{k=1}^{p-1} ( a_{k} \tilde{\varphi}_{c}^{k} (\bm{s}) + b_{k} \tilde{\varphi}_{s}^{k} (\bm{s}),
\label{Filter Para combination}
\end{equation}
where $\tilde{\phi}$ is the filter, $\tilde{\varphi}_{}^{k}$ are the 1D Fourier basis functions, $k=0,1,\cdot\cdot\cdot,p-1$, $p$ denotes the filter size, $a$ and $b$ are learned expansion coefficients. Nonetheless, the aliasing effect in 2D Fourier series expansion for the rotated cases is significant. Following Xie et al.~\cite{Xie2022Fourier}, we replace high-frequency bases with the mirror functions of low-frequency bases to alleviate aliasing effect in rotated 2D Fourier series expansion. 

By replacing the standard filters in the proposed equivariant convolutions with these parameterized filters, DUN-SRE achieve an arbitrary angular resolution w.r.t. the sampled filter orientations without suffering interpolation artifacts from filter rotation. Moreover, it enables more accurate extraction of fine-grained features exhibiting rotational symmetry.

\section{Experimental Results}
\subsection{Experimental Setup}
\subsubsection{Datasets} 
Two cardiac datasets were used in this work: an in-house acquired cardiac cine dataset and the public OCMR dataset~\cite{chen2020ocmr}. The in-house dataset was mainly used for retrospective studies to demonstrate the effectiveness of the proposed method, while the OCMR dataset was used for generalization and prospective studies. Informed consent was obtained from the imaging subjects in accordance with Institutional Review Board (IRB) policy.

\noindent\textbf{In-House Acquired Cardiac cine Data}: The fully sampled cardiac cine data were collected from 29 healthy volunteers using a 3T scanner (MAGNETOM Trio, Siemens Healthcare, Erlangen, Germany). For each subject, $10\sim13$ short-axis slices were imaged with balanced steady-state free precession (bSSFP) sequence during breath-hold, resulting in a total of 356 slices. The following parameters were used: FOV = $330\times330$ mm, acquisition matrix = $256\times256$, slice thickness = 6 mm, TE/TR = 1.5/3.0 ms. The acquired temporal resolution was 40.0 ms and each data has 25 phases. Data augmentation using rigid transformation-shearing was applied. Finally, we obtained 800 multi-coil cardiac MR data of size $192 \times 192 \times 18$ ($x\times y \times t$) for training, 30 for validation, and 118 for testing.
 
\noindent\textbf{OCMR CINE Data}: The OCMR dataset~\cite{chen2020ocmr} used in this study consists of two parts: 37 fully sampled cases for evaluating the generalizability across different scanners and acquisition settings, and 26 prospectively undersampled cases for the prospective study. All samples were collected on a 3T Siemens MAGNETOM Prisma scanner with FOV = $800 \times 300$ mm, acquisition matrix = $384 \times 144$, slice thickness = 8 mm, TE/TR = 1.05/38.4 ms, temporal resolution = 38.4 ms. The 9-fold undersampled data with pseudo-random trajectory were collected in the real-time mode under free-breathing conditions with 65 frames for each prospective data.

\subsubsection{Implementation Details and Evaluation Metrics}
The proposed method was implemented in PyTorch (v1.9.0) and trained for 50 epochs on an Nvidia A100 GPU. The ADAM optimizer was utilized, with an initial learning rate of 0.001, which was adjusted using an exponential decay rate of 0.95. Training and testing were performed on retrospectively undersampled cine data using Cartesian VISTA masks~\cite{ahmad2015VSITA} with different acceleration factors $R$. The coil sensitivity maps were calculated for multicoil data from the fully-filled, time-averaged central k-space data using the ESPIRiT algorithm~\cite{uecker2014espirit}. In proposed SREC, the equivariant number $N_{G}$ is set as 4, \ie, the network is rotation equivariant with $\pi/2$ rotations. The number of equivariant convolution channels is set as $1/4$ to the original regular convolutional network, so that the memory usage is similar to the original network. The number of unrolling iterations is empirically set to 10.

Both visual comparisons and quantitative metrics were used to evaluate reconstruction performance. Quantitative results were evaluated using Peak Signal-to-Noise Ratio (PSNR) on complex-valued images, along with Structural Similarity Index (SSIM)~\cite{wang2004SSIM} and High-Frequency Error Norm (HFEN)~\cite{ravishankar2010HFEN} on magnitude images. HFEN is used to quantify the quality of edges and fine features reconstruction. Higher PSNR, higher SSIM, and lower HFEN indicate better image fidelity. 

\begin{table}[!t]
  \caption{Detailed configurations of model variants for ablation study.\\}\label{Ablation study}
  \centering
  \footnotesize
      \begin{tabular}{lllc}
        \hline \hline 
        \vspace{-7pt}\\ 
        \vspace{2pt} Model & Proximal Net & DC Net & Filter Parametrization \\
        \hline  
        \vspace{-7pt}\\ 
        Baseline & VCNN & VCNN & \xmark \\
        2D-ECNN & 2D ECNN & 2D ECNN & \xmark \\
        SREC-Prox & SREC & VCNN & \xmark \\
        SREC-ProxDC & SREC & SREC & \xmark \\
        \vspace{2pt} DUN-SRE & SREC & SREC & \cmark \\
        \hline \hline 
    \end{tabular}
\end{table}

\subsection{Ablation Studies}
\label{sec:ablation_studies}
To systematically evaluate the effectiveness of each module of the proposed DUN-SRE, we conduct a series of ablation studies by progressively modifying the Proximal Net, DC Net, and the filter parameterization strategy. The detailed configurations of each model variant are summarized in Tab.~\ref{Ablation study}. Vanilla CNN (VCNN) refers to regular convolutions. 2D ECNN are included to assess the effectiveness of directly applying existing 2D equivariant convolutions~\cite{Xie2022Fourier} to dynamic image reconstruction tasks. We conduct comprehensive 3-fold cross-validation to rigorously assess model performance. 

\begin{figure*}[!t]
\centering
\includegraphics[width=0.95\textwidth]{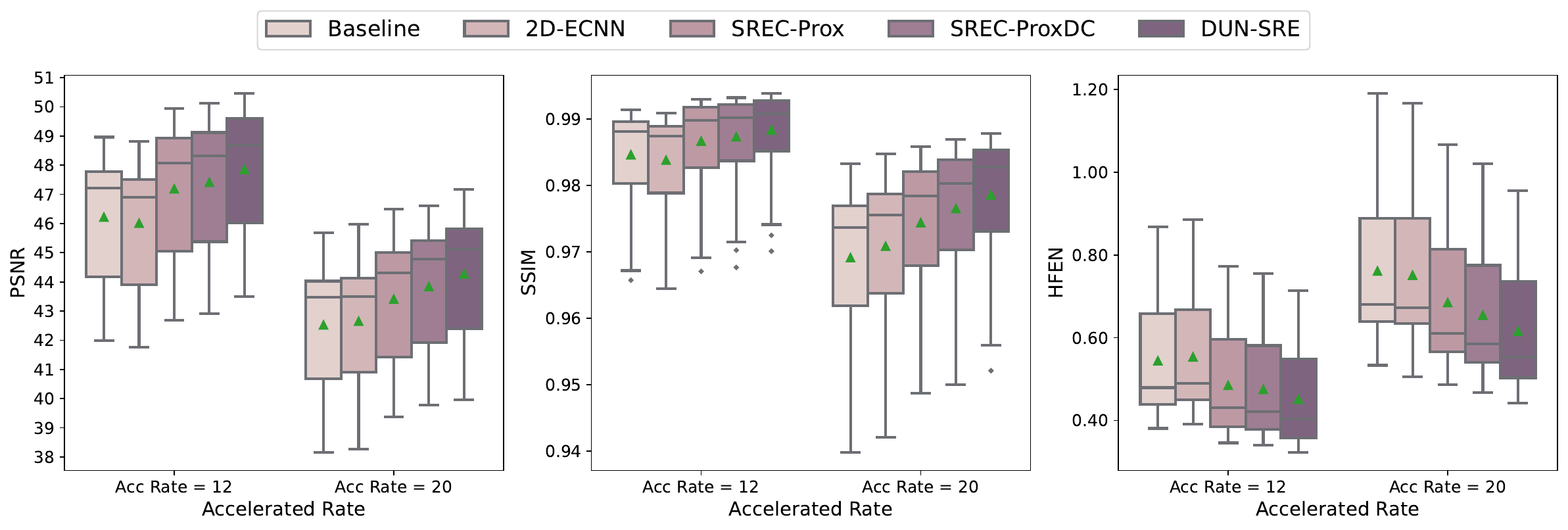}
\caption{Ablation study of the proposed method DUN-SRE on the in-house acquired cardiac cine dataset. Quantitative results of model variants evaluated under two acceleration rates (12× and 20×) across three metrics. The green triangle indicates the mean value.}
\label{Ablation Boxplot}
\end{figure*}

\begin{figure}[!t]
\centering
\includegraphics[width=0.48\textwidth]{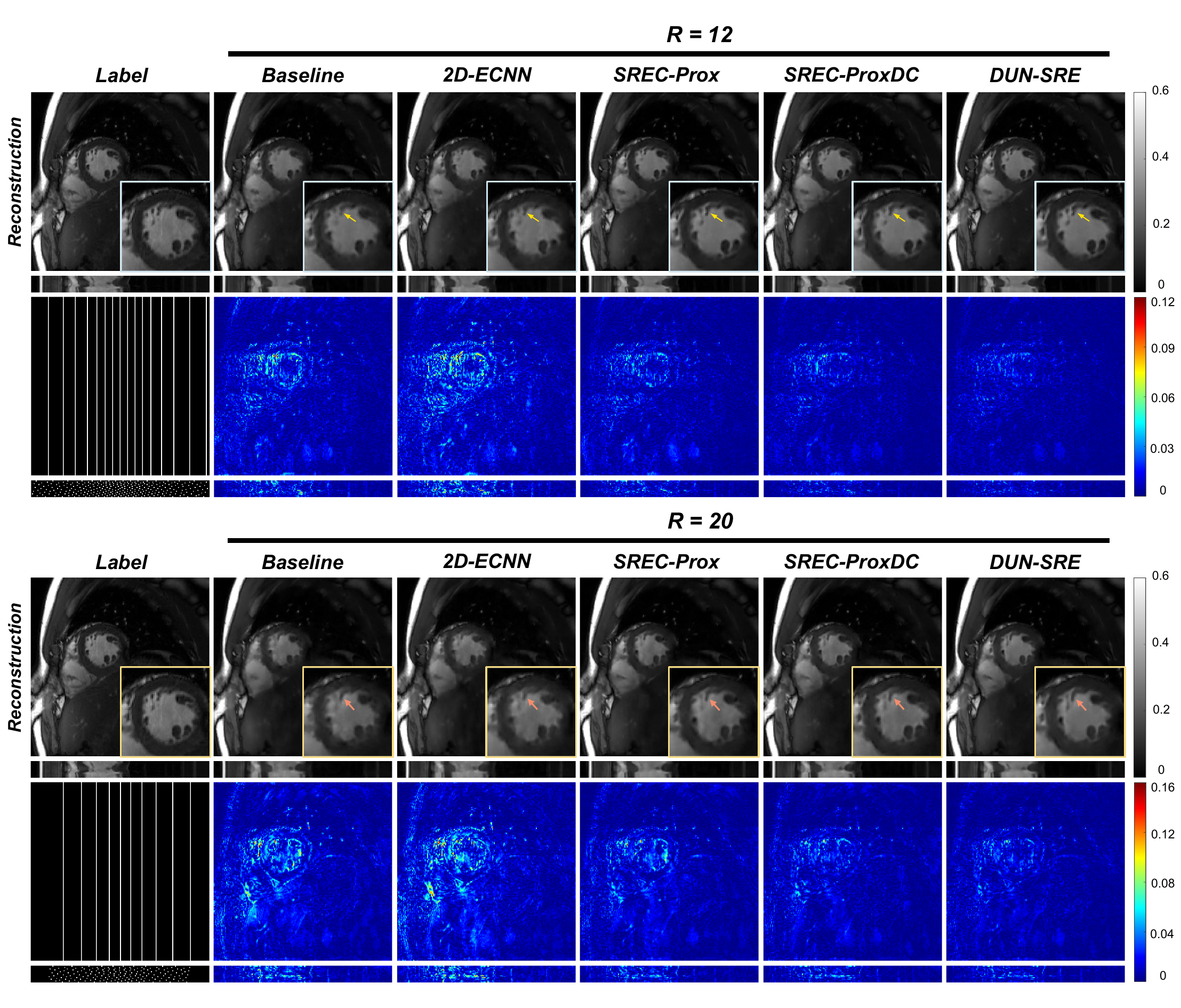}
\caption{Qualitative results of ablations under 12× and 20× acceleration on a sample from the in-house cardiac cine dataset. The first row shows the reconstructed images with zoomed-in views on myocardial regions, and the third row shows the corresponding error maps with the applied sampling mask.}
\label{Ablation}
\end{figure}

First, we evaluated the performance of 2D-ECNN. As shown in Fig. \ref{Ablation Boxplot}, under 12× acceleration, 2D-ECNN performs worse than the baseline across all three metrics. Although a slight improvement is observed at 20× acceleration, the gain is marginal and inconsistent. Visual results in Fig. \ref{Ablation} further show that 2D-ECNN introduces more pronounced artifacts in both spatial and temporal views. These results collectively suggest that directly applying existing 2D ECNN to dynamic reconstruction is ineffective in exploiting symmetry priors and may even degrade reconstruction quality.

SREC-Prox achieves substantial improvements over the baseline as shown in Fig. \ref{Ablation Boxplot}. Moreover, the qualitative results in Figs. \ref{Ablation} reveal that SREC-Prox reconstructs clearer structural details with reduced residual errors, particularly along the myocardial boundaries and papillary muscles of the left ventricle, where rich rotational symmetries are present. These results suggest that SREC establishes a stable framework to achieve rotational equivariance along the temporal dimension, thereby effectively leveraging spatiotemporal rotation symmetry priors to enhance reconstruction quality.

We further evaluate the effectiveness of incorporating the SREC-based equivariant DC network. As shown in the boxplot, SREC-ProxDC demonstrates notable performance gains compared to SREC-Prox. Moreover, the qualitative results in Fig.~\ref{Ablation} show that SREC-ProxDC achieves further improvements in regions characterized by rotation symmetry. These findings indicate that the equivariant DC network more effectively exploits spatiotemporal rotation symmetry priors. 

Finally, DUN-SRE consistently demonstrates superior performance across all quantitative evaluations. In particular, as shown in Fig.~\ref{Ablation}, the zoomed-in regions of DUN-SRE demonstrate clearer reconstruction of fine anatomical structures compared to SREC-ProxDC. It can be observed that the proposed filter parameterization strategy enables high-precision representation of convolutional kernels under rotation, leading to better preservation of symmetry priors and more accurate reconstruction of fine details.

\subsection{Comparisons to State-of-the-Art Models} 
The proposed model is compared with a range of methods. L+S~\cite{otazo2015LS}, a conventional compressed sensing method with low-rank and sparse priors was chosen as the baseline. We also compared DUN-SRE with four deep learning models. 
3D MoDL is a dynamic version of MoDL~\cite{aggarwal2018modl} that employs 3D CNNs to learn noise and aliasing patterns resulting from undersampling, which adopts a 4-layer architecture with filter numbers of 64-64-64-2. In DL-ESPIRiT~\cite{sandino2021DLESPIRIT}, the number of iterations is set to 10, with each iteration comprising three layers with 46-46-2 filters. L+S Net~\cite{huang2021LSNET} combines sparsity prior with a low-rank layer to form a learned soft threshold for rank reduction. It is implemented with 10 iterative sparse prior blocks, each consisting of 3 convolutional layers with 32-32-2 filters, respectively. In unrolled MCMR~\cite{pan2024MCMR} (referred as U-MCMR), we randomly cropped 4 continuous temporal frames out of the 18 temporal frames and 10 unrolled iterations. All deep learning models, including the proposed one, were implemented with approximately 340k parameters and used convolutional filters with a kernel size of 3 to ensure fair comparisons. A consistent training strategy is adopted across all methods, where models are trained and tested on data with varying acceleration rates ($R$ = 8, 12, 16, 20 and 24).

The quantitative comparison results are summarized in Tab.~\ref{Table Comparison Study}. The proposed DUN-SRE consistently outperforms both conventional and learning-based approaches across all evaluated undersampling rates and metrics. The advantage of DUN-SRE becomes more pronounced at higher acceleration factors ($R = 20$ and 24), where it demonstrates increased robustness to severe information loss. The dual improvement in SSIM (global similarity) and HFEN (local edge preservation) metrics confirms the success of DUN-SRE in balancing the often-competing demands of anatomical consistency and high-frequency recovery.
\begin{figure*}[!t]
\centering
\includegraphics[width=0.9\textwidth]{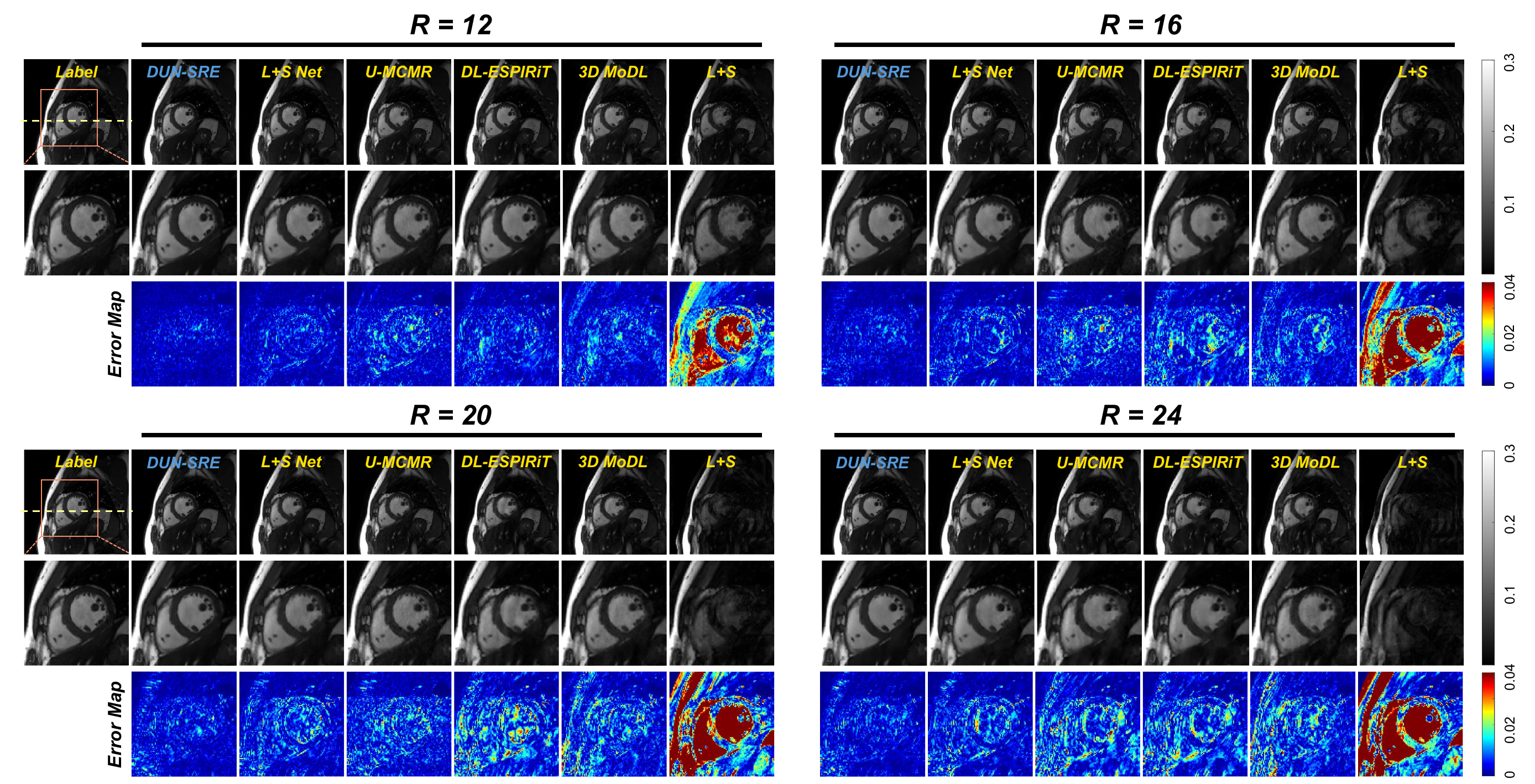}
\caption{Qualitative reconstruction results of comparison study in the spatial domain (x–y) under different acceleration factors. The second row shows the zoom-in region enclosed by the orange box in the reference, and the third row presents the corresponding error maps.}
\label{Comparison 1}
\end{figure*}

\begin{figure}[!t]
\centering
\includegraphics[width=0.48\textwidth]{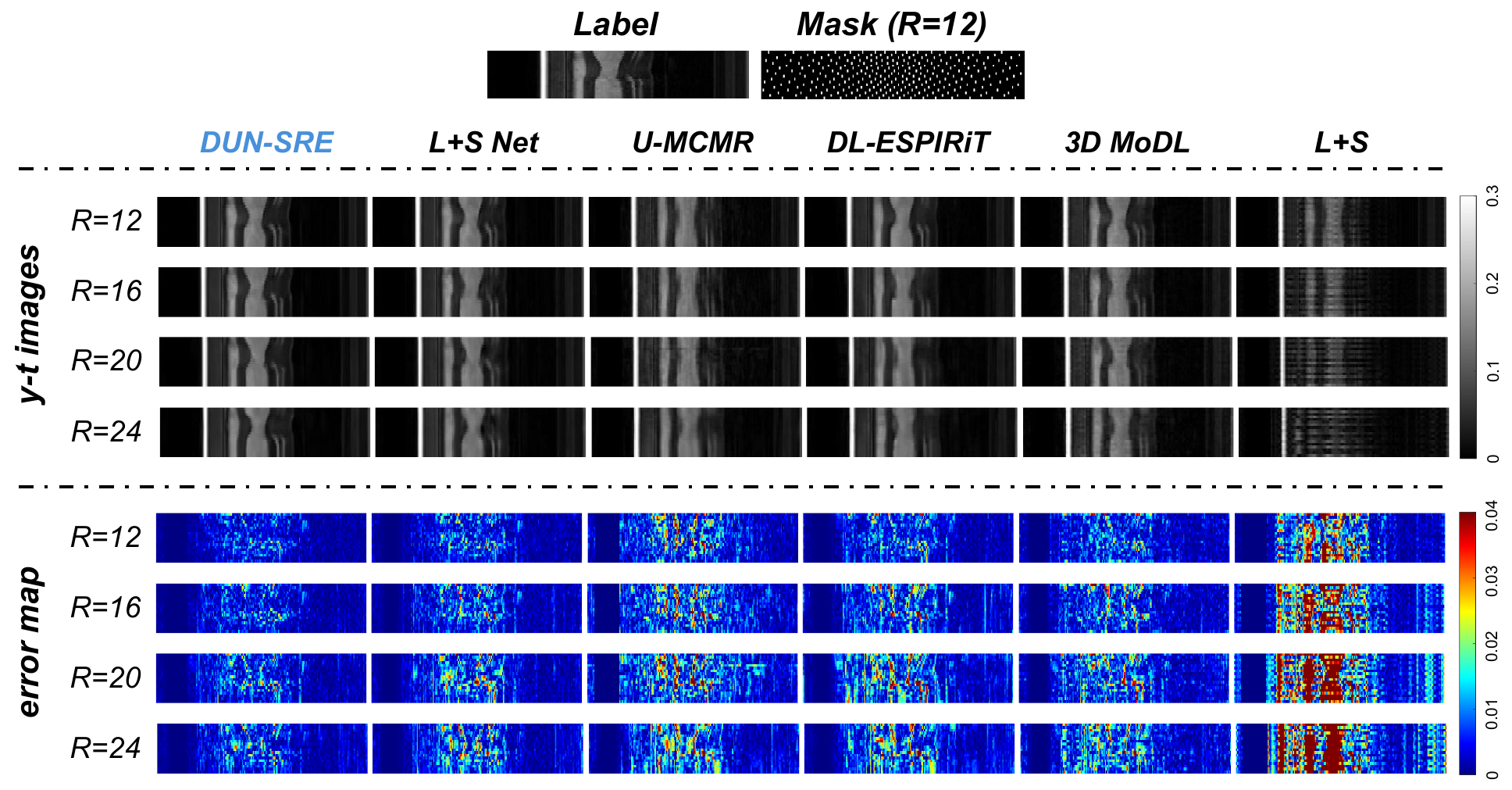}
\caption{Results of comparison study along the temporal dimension (y–t) under different acceleration factors.The selected y-axis, marked by yellow dashed lines in the reference image in Fig. \ref{Comparison 1}. The fully sampled image and an example of the undersampling mask with R = 12 are shown on the top.}
\label{Comparison 2}
\end{figure}

\renewcommand{\arraystretch}{0.95}
\begin{table}[t]
\caption{The average reconstruction results (mean $\pm$ deviation) of all competing methods on the in-house acquired cardiac cine dataset with different acceleration factors of $R=8$, 12, 16, 20, and 24. Best results are marked in bold.}
\label{Table Comparison Study}
\centering
\scriptsize  
\begin{tabular}{>{\raggedright\arraybackslash}p{0.2cm} >{\raggedright\arraybackslash}p{1.8cm} >{\raggedright\arraybackslash}p{1.5cm} >{\raggedright\arraybackslash}p{1.5cm} >{\raggedright\arraybackslash}p{1.5cm}}
\toprule
$R$ &  Methods & PSNR $\uparrow$ & SSIM $\uparrow$ & HFEN $\downarrow$\\
\midrule
   &  L+S~\cite{otazo2015LS}                      & 42.4922$\pm$2.0616  & 0.9719$\pm$0.0129  & 0.7082$\pm$0.1688 \\
   &  3D MoDL~\cite{aggarwal2018modl}             & 46.3945$\pm$1.8059  & 0.9839$\pm$0.0059  & 0.5136$\pm$0.1113 \\
8  &  DL-ESPIRiT~\cite{sandino2021DLESPIRIT}      & 46.9159$\pm$2.0174  & 0.9865$\pm$0.0059  & 0.4998$\pm$0.1218 \\
   &  U-MCMR~\cite{pan2024MCMR}                   & 46.3844$\pm$2.0197  & 0.9865$\pm$0.0053  & 0.5308$\pm$0.1318 \\
   &  L+S Net~\cite{huang2021LSNET}               & 48.3844$\pm$2.0353  & 0.9898$\pm$0.0046  & 0.4416$\pm$0.1108 \\
   &  Our DUN-SRE                                 & \bfseries 49.6794$\pm$2.1122  & \bfseries 0.9915$\pm$0.0045  & \bfseries 0.3710$\pm$0.0958 \\
\midrule
   &  L+S~\cite{otazo2015LS}                      & 37.5190$\pm$2.3518  & 0.9331$\pm$0.0347  & 1.0287$\pm$0.2330 \\
   &  3D MoDL~\cite{aggarwal2018modl}             & 42.4620$\pm$1.6709  & 0.9628$\pm$0.0099  & 0.7082$\pm$0.1424 \\
12 &  DL-ESPIRiT~\cite{sandino2021DLESPIRIT}      & 43.8383$\pm$1.9746  & 0.9765$\pm$0.0090  & 0.6647$\pm$0.1473 \\
   &  U-MCMR~\cite{pan2024MCMR}                   & 43.8173$\pm$2.0293  & 0.9736$\pm$0.0063  & 0.7961$\pm$0.2291 \\
   &  L+S Net~\cite{huang2021LSNET}               & 46.2187$\pm$2.0092  & 0.9849$\pm$0.0060  & 0.5636$\pm$0.1359 \\
   &  Our DUN-SRE                                 & \bfseries 47.8550$\pm$2.0916   & \bfseries 0.9883$\pm$0.0057   & \bfseries 0.4515$\pm$0.1143 \\
\midrule
   &  L+S~\cite{otazo2015LS}                      & 33.9077$\pm$2.0584  & 0.8720$\pm$0.0493  & 1.4120$\pm$0.2657 \\
   &  3D MoDL~\cite{aggarwal2018modl}             & 41.1211$\pm$1.8775  & 0.9607$\pm$0.0118  & 0.8297$\pm$0.1564 \\
16 & DL-ESPIRiT~\cite{sandino2021DLESPIRIT}       & 41.2129$\pm$1.8468  & 0.9596$\pm$0.0129  & 0.8865$\pm$0.1775 \\
   &  U-MCMR~\cite{pan2024MCMR}                   & 41.0131$\pm$2.1806  & 0.9529$\pm$0.0191  & 0.9592$\pm$0.2403 \\
   & L+S Net~\cite{huang2021LSNET}                & 44.5447$\pm$2.0413  & 0.9789$\pm$0.0081  & 0.6609$\pm$0.1585 \\
   &  Our DUN-SRE                                 & \bfseries 45.9255$\pm$2.1549   & \bfseries 0.9837$\pm$0.0080   & \bfseries 0.5356$\pm$0.1302 \\
\midrule
   &  L+S~\cite{otazo2015LS}                      & 30.0754$\pm$1.8259  & 0.7878$\pm$0.0536  & 1.9649$\pm$0.3294 \\
   &  3D MoDL~\cite{aggarwal2018modl}             & 39.2965$\pm$1.5566  & 0.9448$\pm$0.0134  & 0.9721$\pm$0.1733 \\
20 & DL-ESPIRiT~\cite{sandino2021DLESPIRIT}       & 39.4161$\pm$1.8306  & 0.9449$\pm$0.0159  & 1.0324$\pm$0.2030 \\
   &  U-MCMR~\cite{pan2024MCMR}                   & 39.0767$\pm$1.9860  & 0.9303$\pm$0.0247  & 1.0588$\pm$0.2528 \\
   & L+S Net~\cite{huang2021LSNET}                & 43.0857$\pm$2.0256  & 0.9741$\pm$0.0096  & 0.7427$\pm$0.1720 \\
   &  Our DUN-SRE                                 & \bfseries 44.2714$\pm$2.0066   & \bfseries 0.9786$\pm$0.0089   & \bfseries 0.6155$\pm$0.1442 \\
\midrule
   &  L+S~\cite{otazo2015LS}                      & 27.1966$\pm$1.5994  & 0.7122$\pm$0.0511  & 2.4233$\pm$0.3474 \\
   &  3D MoDL~\cite{aggarwal2018modl}             & 37.3473$\pm$1.5662  & 0.9214$\pm$0.0172  & 1.1734$\pm$0.1818 \\
24 & DL-ESPIRiT~\cite{sandino2021DLESPIRIT}       & 38.6397$\pm$1.7230  & 0.9373$\pm$0.0169  & 1.1084$\pm$0.1991 \\
   &  U-MCMR~\cite{pan2024MCMR}                   & 38.6397$\pm$1.7230  & 0.9373$\pm$0.0168  & 1.3084$\pm$0.2090 \\
   &  L+S Net~\cite{huang2021LSNET}               & 41.8881$\pm$1.9159  & 0.9673$\pm$0.0110  & 0.8263$\pm$0.1750 \\
   &  Our DUN-SRE                                 & \bfseries 43.2553$\pm$1.9148   & \bfseries 0.9747$\pm$0.0095   & \bfseries 0.6950$\pm$0.1548 \\
\bottomrule
\end{tabular}   
\end{table}

The visual comparisons to the compared methods together with the error maps are shown in Fig. \ref{Comparison 1} and Fig. \ref{Comparison 2}. In the spatial domain (x–y slices), DUN-SRE consistently outperforms all compared methods in all acceleration factors with fewer aliasing artifacts and less residual error. Notably, DUN-SRE achieves significantly better reconstruction in regions exhibiting rotation symmetry, such as the myocardial boundaries of the left ventricle and the contours of the papillary muscles. Temporal fidelity is evaluated through the y–t slices in Fig.~\ref{Comparison 2}. DUN-SRE accurately recovers cardiac dynamics and motion patterns with reduced residual errors. In contrast, compared methods suffer from temporal blurring (\eg, L+S Net, 3D MoDL) or residual aliasing (\eg, L+S, DL-ESPIRiT), particularly under large acceleration factors.

\begin{figure}[!t]
\centering
\includegraphics[width=0.48\textwidth]{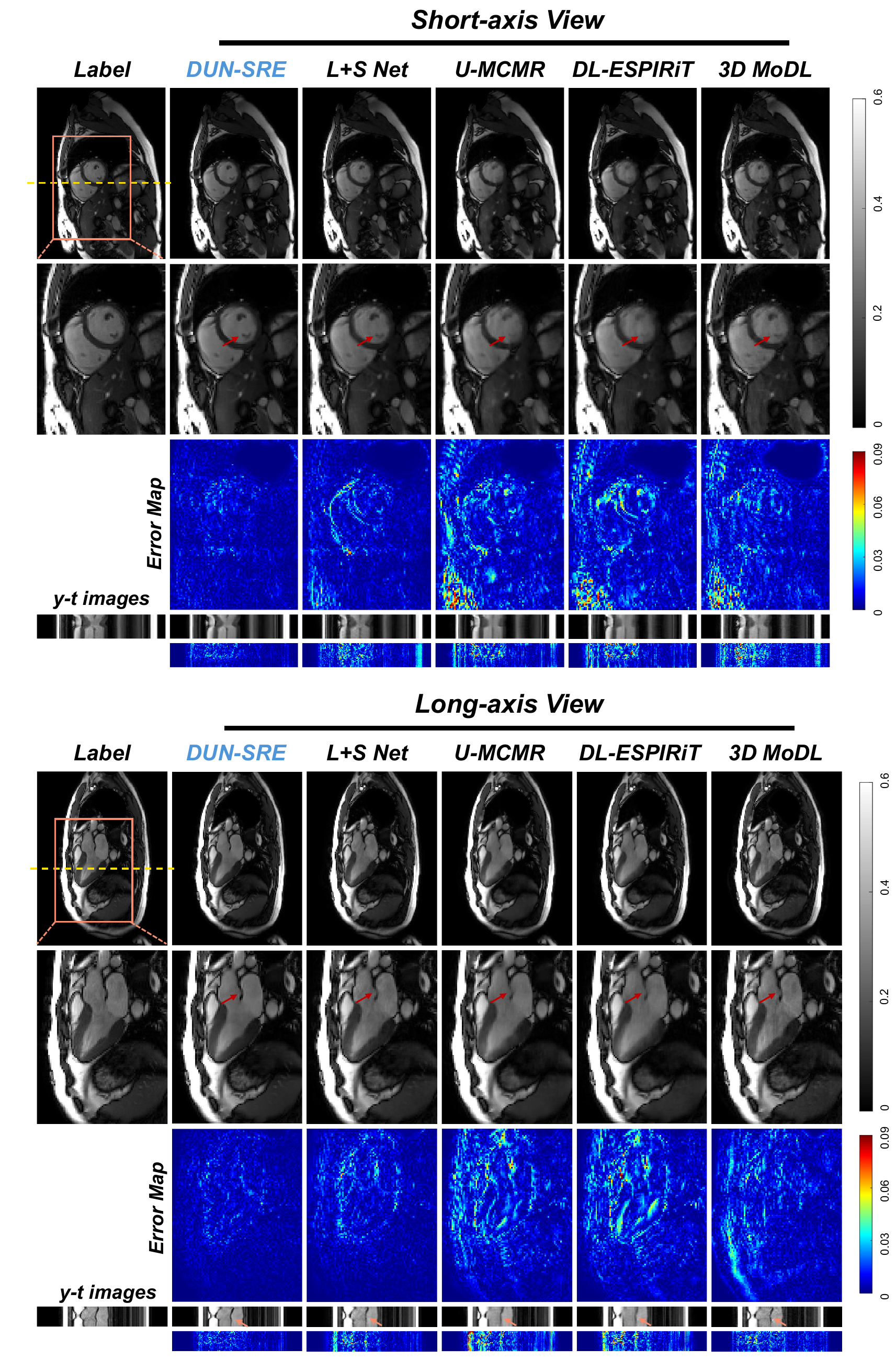}
\caption{Visual comparisons for the generalization study on the OCMR dataset~\cite{chen2020ocmr} under 9× acceleration. The second row shows zoomed-in regions corresponding to the orange boxes in the reference images. The fourth row presents the y–t images extracted along the y-axis marked by the yellow dashed lines. Red arrows highlight differences in anatomical structure details.}
\label{Generalization Study}
\end{figure}

\begin{figure}[!t]
\centering
\includegraphics[width=0.48\textwidth]{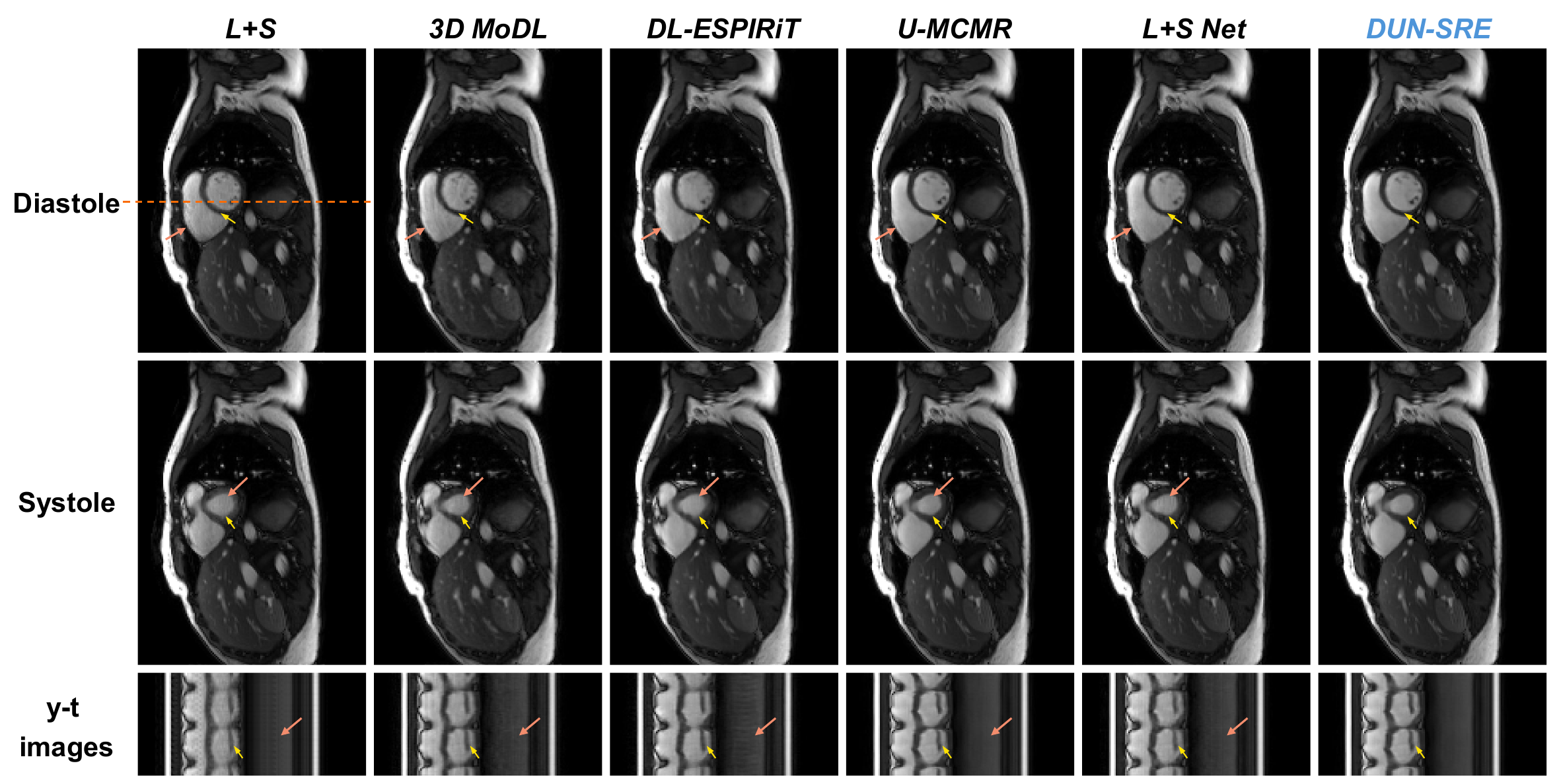}
\caption{Reconstruction results of prospective study on real-time 9× undersampled OCMR dataset~\cite{chen2020ocmr}. The third row presents the y–t images along the orange dashed line in the diastolic frame. Pink arrows indicate blurring artifacts, while yellow arrows highlight differences in anatomical structure details. }
\label{Prospective}
\end{figure}

\subsection{Generalization and Prospective Study}
Finally, we evaluate the models' generalization capabilities on fully sampled OCMR dataset~\cite{chen2020ocmr}, which was acquired with different scanners and imaging protocols. In addition, we evaluate the robustness and practical applicability of the proposed method on real-time prospectively undersampled OCMR dataset~\cite{chen2020ocmr}. Specifically, the 9× undersampled OCMR data were tested using models trained at 12× acceleration, without any transfer learning or fine-tuning. 

For generalization study, Fig.~\ref{Generalization Study} presents reconstruction results for a subject with different views. The long-axis view is not included in the training set. The proposed method demonstrates strong generalization ability, consistently outperforming other compared methods in terms of detail preservation and artifact suppression on previously unseen test data. 

For prospective study, Fig.~\ref{Prospective} presents visual comparisons at both the diastolic and systolic cardiac phases of a subject. DUN-SRE outperforms all the competing methods with less spatial blurring and streaking artifacts. In addition, DUN-SRE achieves superior reconstruction of fine anatomical structures and myocardial boundaries, and successfully restores relevant diagnostic features such as the papillary muscles without blurring. Furthermore, the motion fields generated by DUN-SRE are denser, smoother, and more physiologically meaningful than those generated by other methods. These results demonstrate its strong generalizability and robustness in real-world acquisition scenarios involving domain shifts.

\section{Discussions} 
Experimental results consistently demonstrate that DUN-SRE offers superior performance in recovering anatomical structures that exhibit rotational symmetry, \eg, the myocardial boundaries of the left and right ventricles and the papillary muscles. These regions are clinically significant for diagnosing pathological conditions and assessing cardiac function. The preserved rotational features confirm DUN-SRE's effective learning of spatiotemporal rotation-equivariant representations, directly improving reconstruction fidelity. 

DUN-SRE's superiority stems from its spatiotemporal convolutional architecture, effectively modeling symmetry in space and time. While standard 2D and 3D ECNNs have been shown to preserve spatial symmetry, they are inherently limited in their ability to capture temporal equivariance. As evidenced by the ablation study in Sec.~\ref{sec:ablation_studies}, directly extending 2D ECNN to dynamic image reconstruction leads to suboptimal results, primarily due to the breakdown of group-equivariant representations when temporal convolutions are included in a naïve manner, \ie, 1D convolutions applied along the temporal dimension may disrupt the group structure established by preceding 2D layers, thereby compromising the model’s ability to maintain spatial equivariance over time. Furthermore, volumetric 3D ECNNs, which are generally developed to model global 3D object rotations, are not well-suited for dynamic imaging, where data are structured as 2D frames evolving along a temporal axis. The spatiotemporal structure of dynamic MRI (\ie, 2D+t) fundamentally differs from static volumetric data, making 3D ECNN less effective in modeling temporal consistency and motion-related symmetries.

To leverage the spatiotemporal rotational symmetry, the proposed SREC module employs a novel (2+1)D architecture, wherein spatial and temporal operations are decoupled while maintaining equivariance. This structure allows the network to first extract stable rotation-equivariant features from individual frames using 2D group-equivariant convolutions, followed by temporal modeling that captures structural redundancy across time. Such a design enables the model to leverage spatiotemporal symmetry in a structured and principled manner to better extract the characteristics of dynamic MRI.

The proposed DUN-SRE also benefits from improved weight-sharing efficiency. In ECNN, convolutional filters are inherently shared across transformed versions of patterns through group actions, significantly reducing the number of learnable parameters. By extending this mechanism to the temporal domain, SREC enables filter sharing not only within individual frames but also across temporally aligned frames, effectively increasing parameter efficiency by a factor proportional to the frame length. This leads to a more compact and data-efficient representation, which is particularly advantageous in scenarios with limited training data, and contributes to the model’s generalization ability.


Lastly, DUN-SRE's strong generalization stems from its rotation symmetry prior, which is more fundamental than data-driven adaptations. This is reflected in our generalization and prospective experiments, where DUN-SRE consistently achieves strong performance without the need for fine-tuning. Moreover, the spatiotemporal symmetry modeling strategy adopted in this work has the potential to benefit other dynamic imaging and video-based signal processing tasks, suggesting broader applicability beyond cardiac MRI.

\section{Conclusion}
In this work, we propose DUN-SRE, a deep unrolling nework for dynamic MRI reconstruction that fully exploits rotational symmetry priors across both spatial and temporal dimensions. By embedding rotation equivariance throughout the entire unrolling optimization process, the proposed method enables more effective modeling of dynamic image structures. We compared DUN-SRE with state-of-the-art models on cardiac cine MRI datasets. Experimental results demonstrate its superior performance, particularly in preserving fine anatomical structures and regions characterized by rotational symmetry. DUN-SRE's principled design enables robust performance in dynamic MRI reconstruction and could generalize well to other spatiotemporal tasks like video restoration and motion-resolved imaging. 


%

\balance
\bibliographystyle{IEEEtran}
\bibliography{IEEEabrv,reference}

\begin{thebibliography}{10}
\providecommand{\url}[1]{#1}
\csname url@samestyle\endcsname
\providecommand{\newblock}{\relax}
\providecommand{\bibinfo}[2]{#2}
\providecommand{\BIBentrySTDinterwordspacing}{\spaceskip=0pt\relax}
\providecommand{\BIBentryALTinterwordstretchfactor}{4}
\providecommand{\BIBentryALTinterwordspacing}{\spaceskip=\fontdimen2\font plus
\BIBentryALTinterwordstretchfactor\fontdimen3\font minus \fontdimen4\font\relax}
\providecommand{\BIBforeignlanguage}[2]{{%
\expandafter\ifx\csname l@#1\endcsname\relax
\typeout{** WARNING: IEEEtran.bst: No hyphenation pattern has been}%
\typeout{** loaded for the language `#1'. Using the pattern for}%
\typeout{** the default language instead.}%
\else
\language=\csname l@#1\endcsname
\fi
#2}}
\providecommand{\BIBdecl}{\relax}
\BIBdecl

\bibitem{keall2022integrated}
P.~J. Keall, C.~Brighi, C.~Glide-Hurst, G.~Liney, P.~Z. Liu, S.~Lydiard, C.~Paganelli, T.~Pham, S.~Shan, A.~C. Tree \emph{et~al.}, ``Integrated {MRI}-guided radiotherapy—opportunities and challenges,'' \emph{Nature Reviews Clinical Oncology}, vol.~19, no.~7, pp. 458--470, 2022.

\bibitem{wang2024cine}
Y.-R. Wang, K.~Yang, Y.~Wen, P.~Wang, Y.~Hu, Y.~Lai, Y.~Wang, K.~Zhao, S.~Tang, A.~Zhang \emph{et~al.}, ``Screening and diagnosis of cardiovascular disease using artificial intelligence-enabled cardiac magnetic resonance imaging,'' \emph{Nature Medicine}, vol.~30, no.~5, pp. 1471--1480, 2024.

\bibitem{Lustig2007SparseMR}
M.~Lustig, D.~Donoho, and J.~M. Pauly, ``Sparse {MRI}: The application of compressed sensing for rapid {MR} imaging,'' \emph{Magnetic Resonance in Medicine: An Official Journal of the International Society for Magnetic Resonance in Medicine}, vol.~58, no.~6, pp. 1182--1195, 2007.

\bibitem{fessler2010model}
J.~A. Fessler, ``Model-based image reconstruction for {MRI},'' \emph{IEEE signal processing magazine}, vol.~27, no.~4, pp. 81--89, 2010.

\bibitem{liu2020rare}
J.~Liu, Y.~Sun, C.~Eldeniz, W.~Gan, H.~An, and U.~S. Kamilov, ``{RARE}: Image reconstruction using deep priors learned without groundtruth,'' \emph{IEEE Journal of Selected Topics in Signal Processing}, vol.~14, no.~6, pp. 1088--1099, 2020.

\bibitem{fessler2020optimization}
J.~A. Fessler, ``Optimization methods for magnetic resonance image reconstruction: Key models and optimization algorithms,'' \emph{IEEE signal processing magazine}, vol.~37, no.~1, pp. 33--40, 2020.

\bibitem{zhao2012sparsity}
B.~Zhao, J.~P. Haldar, A.~G. Christodoulou, and Z.-P. Liang, ``Image reconstruction from highly undersampled (k, t)-space data with joint partial separability and sparsity constraints,'' \emph{IEEE transactions on medical imaging}, vol.~31, no.~9, pp. 1809--1820, 2012.

\bibitem{haldar2013rank}
J.~P. Haldar, ``Low-rank modeling of local $ k $-space neighborhoods ({LORAKS}) for constrained {MRI},'' \emph{IEEE transactions on medical imaging}, vol.~33, no.~3, pp. 668--681, 2013.

\bibitem{He2016lowrk}
J.~He, Q.~Liu, A.~G. Christodoulou, C.~Ma, F.~Lam, and Z.-P. Liang, ``Accelerated high-dimensional {MR} imaging with sparse sampling using low-rank tensors,'' \emph{IEEE transactions on medical imaging}, vol.~35, no.~9, pp. 2119--2129, 2016.

\bibitem{hinton2006reducing}
G.~E. Hinton and R.~R. Salakhutdinov, ``Reducing the dimensionality of data with neural networks,'' \emph{science}, vol. 313, no. 5786, pp. 504--507, 2006.

\bibitem{lecun2015deep}
Y.~LeCun, Y.~Bengio, and G.~Hinton, ``Deep learning,'' \emph{nature}, vol. 521, no. 7553, pp. 436--444, 2015.

\bibitem{yu2018aihealth}
K.-H. Yu, A.~L. Beam, and I.~S. Kohane, ``Artificial intelligence in healthcare,'' \emph{Nature biomedical engineering}, vol.~2, no.~10, pp. 719--731, 2018.

\bibitem{Liang2020deep}
D.~Liang, J.~Cheng, Z.~Ke, and L.~Ying, ``Deep magnetic resonance image reconstruction: Inverse problems meet neural networks,'' \emph{IEEE Signal Processing Magazine}, vol.~37, no.~1, pp. 141--151, 2020.

\bibitem{hosseini2020RCNMR}
S.~A.~H. Hosseini, B.~Yaman, S.~Moeller, M.~Hong, and M.~Ak{\c{c}}akaya, ``Dense recurrent neural networks for accelerated {MRI}: History-cognizant unrolling of optimization algorithms,'' \emph{IEEE Journal of Selected Topics in Signal Processing}, vol.~14, no.~6, pp. 1280--1291, 2020.

\bibitem{cha2020geometric}
E.~Cha, G.~Oh, and J.~C. Ye, ``Geometric approaches to increase the expressivity of deep neural networks for {MR} reconstruction,'' \emph{IEEE Journal of Selected Topics in Signal Processing}, vol.~14, no.~6, pp. 1292--1305, 2020.

\bibitem{zhang2022unfoldmr}
J.~Zhang, Z.~Zhang, J.~Xie, and Y.~Zhang, ``High-throughput deep unfolding network for compressive sensing {MRI},'' \emph{IEEE Journal of Selected Topics in Signal Processing}, vol.~16, no.~4, pp. 750--761, 2022.

\bibitem{chen2022aimr}
Y.~Chen, C.-B. Sch{\"o}nlieb, P.~Li{\`o}, T.~Leiner, P.~L. Dragotti, G.~Wang, D.~Rueckert, D.~Firmin, and G.~Yang, ``{AI}-based reconstruction for fast {MRI—A} systematic review and meta-analysis,'' \emph{Proceedings of the IEEE}, vol. 110, no.~2, pp. 224--245, 2022.

\bibitem{qin2018CRNN}
C.~Qin, J.~Schlemper, J.~Caballero, A.~N. Price, J.~V. Hajnal, and D.~Rueckert, ``Convolutional recurrent neural networks for dynamic {MR} image reconstruction,'' \emph{IEEE transactions on medical imaging}, vol.~38, no.~1, pp. 280--290, 2018.

\bibitem{kustner2020CINENet}
T.~K{\"u}stner, N.~Fuin, K.~Hammernik, A.~Bustin, H.~Qi, R.~Hajhosseiny, P.~G. Masci, R.~Neji, D.~Rueckert, R.~M. Botnar \emph{et~al.}, ``Cinenet: deep learning-based 3d cardiac {CINE MRI} reconstruction with multi-coil complex-valued 4d spatio-temporal convolutions,'' \emph{Scientific reports}, vol.~10, no.~1, p. 13710, 2020.

\bibitem{aggarwal2020jmodl}
H.~K. Aggarwal and M.~Jacob, ``J-modl: Joint model-based deep learning for optimized sampling and reconstruction,'' \emph{IEEE journal of selected topics in signal processing}, vol.~14, no.~6, pp. 1151--1162, 2020.

\bibitem{souza2020EnMRprior}
R.~Souza, Y.~Beauferris, W.~Loos, R.~M. Lebel, and R.~Frayne, ``Enhanced deep-learning-based magnetic resonance image reconstruction by leveraging prior subject-specific brain imaging: Proof-of-concept using a cohort of presumed normal subjects,'' \emph{IEEE Journal of Selected Topics in Signal Processing}, vol.~14, no.~6, pp. 1126--1136, 2020.

\bibitem{martinini2022recovermr}
F.~Martinini, M.~Mangia, A.~Marchioni, R.~Rovatti, and G.~Setti, ``A deep learning method for optimal undersampling patterns and image recovery for {MRI} exploiting losses and projections,'' \emph{IEEE Journal of Selected Topics in Signal Processing}, vol.~16, no.~4, pp. 713--724, 2022.

\bibitem{xie2022puert}
J.~Xie, J.~Zhang, Y.~Zhang, and X.~Ji, ``{PUERT}: Probabilistic under-sampling and explicable reconstruction network for {CS-MRI},'' \emph{IEEE Journal of Selected Topics in Signal Processing}, vol.~16, no.~4, pp. 737--749, 2022.

\bibitem{Szegedy2015CNN}
C.~Szegedy, W.~Liu, Y.~Jia, P.~Sermanet, S.~Reed, D.~Anguelov, D.~Erhan, V.~Vanhoucke, and A.~Rabinovich, ``Going deeper with convolutions,'' in \emph{Proceedings of the IEEE conference on computer vision and pattern recognition}, 2015, pp. 1--9.

\bibitem{Cohen2016GCNN}
T.~Cohen and M.~Welling, ``Group equivariant convolutional networks,'' in \emph{International conference on machine learning}.\hskip 1em plus 0.5em minus 0.4em\relax PMLR, 2016, pp. 2990--2999.

\bibitem{Graham2020Dense}
S.~Graham, D.~Epstein, and N.~Rajpoot, ``Dense steerable filter {CNN}s for exploiting rotational symmetry in histology images,'' \emph{IEEE Transactions on Medical Imaging}, vol.~39, no.~12, pp. 4124--4136, 2020.

\bibitem{lafarge2021medical}
M.~W. Lafarge, E.~J. Bekkers, J.~P. Pluim, R.~Duits, and M.~Veta, ``Roto-translation equivariant convolutional networks: Application to histopathology image analysis,'' \emph{Medical Image Analysis}, vol.~68, p. 101849, 2021.

\bibitem{weiler2019general}
M.~Weiler and G.~Cesa, ``General e (2)-equivariant steerable {CNN}s,'' \emph{Advances in neural information processing systems}, vol.~32, 2019.

\bibitem{Wang2021CT}
H.~Wang, Y.~Li, N.~He, K.~Ma, D.~Meng, and Y.~Zheng, ``Dicdnet: deep interpretable convolutional dictionary network for metal artifact reduction in {CT} images,'' \emph{IEEE Transactions on Medical Imaging}, vol.~41, no.~4, pp. 869--880, 2021.

\bibitem{Xie2022Fourier}
Q.~Xie, Q.~Zhao, Z.~Xu, and D.~Meng, ``Fourier series expansion based filter parametrization for equivariant convolutions,'' \emph{IEEE Transactions on Pattern Analysis and Machine Intelligence}, vol.~45, no.~4, pp. 4537--4551, 2022.

\bibitem{Weiler2018Steerable}
M.~Weiler, F.~A. Hamprecht, and M.~Storath, ``Learning steerable filters for rotation equivariant {CNN}s,'' in \emph{Proceedings of the IEEE Conference on Computer Vision and Pattern Recognition}, 2018, pp. 849--858.

\bibitem{kalogeropoulos2024scale}
I.~Kalogeropoulos, G.~Bouritsas, and Y.~Panagakis, ``Scale equivariant graph metanetworks,'' \emph{Advances in neural information processing systems}, vol.~37, pp. 106\,800--106\,840, 2024.

\bibitem{gerken2023geometric}
J.~E. Gerken, J.~Aronsson, O.~Carlsson, H.~Linander, F.~Ohlsson, C.~Petersson, and D.~Persson, ``Geometric deep learning and equivariant neural networks,'' \emph{Artificial Intelligence Review}, vol.~56, no.~12, pp. 14\,605--14\,662, 2023.

\bibitem{Chen2023unsupervised}
D.~Chen, M.~Davies, M.~J. Ehrhardt, C.-B. Sch{\"o}nlieb, F.~Sherry, and J.~Tachella, ``Imaging with equivariant deep learning: From unrolled network design to fully unsupervised learning,'' \emph{IEEE Signal Processing Magazine}, vol.~40, no.~1, pp. 134--147, 2023.

\bibitem{zhu2024sre}
Y.~Zhu, J.~Cheng, Z.-X. Cui, J.~Ren, C.~Wang, and D.~Liang, ``{SRE-CNN}: A spatiotemporal rotation-equivariant {CNN} for cardiac cine {MR} imaging,'' in \emph{International Conference on Medical Image Computing and Computer-Assisted Intervention}.\hskip 1em plus 0.5em minus 0.4em\relax Springer, 2024, pp. 679--689.

\bibitem{chen2020ocmr}
C.~Chen, Y.~Liu, P.~Schniter, M.~Tong, K.~Zareba, O.~Simonetti, L.~Potter, and R.~Ahmad, ``Ocmr (v1. 0)--open-access multi-coil k-space dataset for cardiovascular magnetic resonance imaging,'' \emph{arXiv preprint arXiv:2008.03410}, 2020.

\bibitem{shen2020pdo}
Z.~Shen, L.~He, Z.~Lin, and J.~Ma, ``Pdo-econvs: Partial differential operator based equivariant convolutions,'' in \emph{International Conference on Machine Learning}.\hskip 1em plus 0.5em minus 0.4em\relax PMLR, 2020, pp. 8697--8706.

\bibitem{cesa2022En}
G.~Cesa, L.~Lang, and M.~Weiler, ``A program to build e (n)-equivariant steerable {CNN}s,'' in \emph{International conference on learning representations}, 2022.

\bibitem{hoogeboom2018hexaconv}
E.~Hoogeboom, J.~W. Peters, T.~S. Cohen, and M.~Welling, ``Hexaconv,'' in \emph{International Conference on Learning Representations}, 2018.

\bibitem{worrall2017harmonic}
D.~E. Worrall, S.~J. Garbin, D.~Turmukhambetov, and G.~J. Brostow, ``Harmonic networks: Deep translation and rotation equivariance,'' in \emph{Proceedings of the IEEE conference on computer vision and pattern recognition}, 2017, pp. 5028--5037.

\bibitem{worrall2018cubenet}
D.~Worrall and G.~Brostow, ``Cubenet: Equivariance to 3d rotation and translation,'' in \emph{Proceedings of the European Conference on Computer Vision (ECCV)}, 2018, pp. 567--584.

\bibitem{cohen2018spherical}
T.~S. Cohen, M.~Geiger, J.~K{\"o}hler, and M.~Welling, ``Spherical {CNN}s,'' in \emph{International Conference on Learning Representations}, 2018.

\bibitem{yang2018admm}
Y.~Yang, J.~Sun, H.~Li, and Z.~Xu, ``{ADMM}-csnet: A deep learning approach for image compressive sensing,'' \emph{IEEE transactions on pattern analysis and machine intelligence}, vol.~42, no.~3, pp. 521--538, 2018.

\bibitem{zhang2018ista}
J.~Zhang and B.~Ghanem, ``{ISTA}-net: Interpretable optimization-inspired deep network for image compressive sensing,'' in \emph{Proceedings of the IEEE conference on computer vision and pattern recognition}, 2018, pp. 1828--1837.

\bibitem{hammernik2018VN}
K.~Hammernik, T.~Klatzer, E.~Kobler, M.~P. Recht, D.~K. Sodickson, T.~Pock, and F.~Knoll, ``Learning a variational network for reconstruction of accelerated mri data,'' \emph{Magnetic resonance in medicine}, vol.~79, no.~6, pp. 3055--3071, 2018.

\bibitem{zheng2019deepDC}
H.~Zheng, F.~Fang, and G.~Zhang, ``Cascaded dilated dense network with two-step data consistency for {MRI} reconstruction,'' \emph{Advances in Neural Information Processing Systems}, vol.~32, 2019.

\bibitem{gungor2022transms}
A.~G{\"u}ng{\"o}r, B.~Askin, D.~A. Soydan, E.~U. Saritas, C.~B. Top, and T.~{\c{C}}ukur, ``Transms: Transformers for super-resolution calibration in magnetic particle imaging,'' \emph{IEEE Transactions on Medical Imaging}, vol.~41, no.~12, pp. 3562--3574, 2022.

\bibitem{liu2023DCprior}
J.~Liu, C.~Qin, and M.~Yaghoobi, ``High-fidelity {MRI} reconstruction using adaptive spatial attention selection and deep data consistency prior,'' \emph{IEEE Transactions on Computational Imaging}, vol.~9, pp. 298--313, 2023.

\bibitem{gungor2023deq}
A.~G{\"u}ng{\"o}r, B.~Askin, D.~A. Soydan, C.~B. Top, E.~U. Saritas, and T.~{\c{C}}ukur, ``{DEQ-MPI}: A deep equilibrium reconstruction with learned consistency for magnetic particle imaging,'' \emph{IEEE Transactions on Medical Imaging}, vol.~43, no.~1, pp. 321--334, 2023.

\bibitem{brown2014magnetic}
R.~W. Brown, Y.-C.~N. Cheng, E.~M. Haacke, M.~R. Thompson, and R.~Venkatesan, \emph{Magnetic resonance imaging: physical principles and sequence design}.\hskip 1em plus 0.5em minus 0.4em\relax John Wiley \& Sons, 2014.

\bibitem{ahmad2020plug}
R.~Ahmad, C.~A. Bouman, G.~T. Buzzard, S.~Chan, S.~Liu, E.~T. Reehorst, and P.~Schniter, ``Plug-and-play methods for magnetic resonance imaging: Using denoisers for image recovery,'' \emph{IEEE signal processing magazine}, vol.~37, no.~1, pp. 105--116, 2020.

\bibitem{cheng2021LDC}
J.~Cheng, Z.-X. Cui, W.~Huang, Z.~Ke, L.~Ying, H.~Wang, Y.~Zhu, and D.~Liang, ``Learning data consistency and its application to dynamic {MR} imaging,'' \emph{IEEE transactions on medical imaging}, vol.~40, no.~11, pp. 3140--3153, 2021.

\bibitem{celledoni2021inverse}
E.~Celledoni, M.~J. Ehrhardt, C.~Etmann, B.~Owren, C.-B. Sch{\"o}nlieb, and F.~Sherry, ``Equivariant neural networks for inverse problems,'' \emph{Inverse Problems}, vol.~37, no.~8, p. 085006, 2021.

\bibitem{gunel2022scale}
B.~Gunel, A.~Sahiner, A.~D. Desai, A.~S. Chaudhari, S.~Vasanawala, M.~Pilanci, and J.~Pauly, ``Scale-equivariant unrolled neural networks for data-efficient accelerated {MRI} reconstruction,'' in \emph{International Conference on Medical Image Computing and Computer-Assisted Intervention}.\hskip 1em plus 0.5em minus 0.4em\relax Springer, 2022, pp. 737--747.

\bibitem{fu2024proximal}
J.~Fu, Q.~Xie, D.~Meng, and Z.~Xu, ``Rotation equivariant proximal operator for deep unfolding methods in image restoration,'' \emph{IEEE Transactions on Pattern Analysis and Machine Intelligence}, 2024.

\bibitem{parikh2014proximal}
N.~Parikh, S.~Boyd \emph{et~al.}, ``Proximal algorithms,'' \emph{Foundations and trends{\textregistered} in Optimization}, vol.~1, no.~3, pp. 127--239, 2014.

\bibitem{tran2018Conv21D}
D.~Tran, H.~Wang, L.~Torresani, J.~Ray, Y.~LeCun, and M.~Paluri, ``A closer look at spatiotemporal convolutions for action recognition,'' in \emph{Proceedings of the IEEE conference on Computer Vision and Pattern Recognition}, 2018, pp. 6450--6459.

\bibitem{ahmad2015VSITA}
R.~Ahmad, H.~Xue, S.~Giri, Y.~Ding, J.~Craft, and O.~P. Simonetti, ``Variable density incoherent spatiotemporal acquisition ({VISTA}) for highly accelerated cardiac mri,'' \emph{Magnetic resonance in medicine}, vol.~74, no.~5, pp. 1266--1278, 2015.

\bibitem{uecker2014espirit}
M.~Uecker, P.~Lai, M.~J. Murphy, P.~Virtue, M.~Elad, J.~M. Pauly, S.~S. Vasanawala, and M.~Lustig, ``Espirit—an eigenvalue approach to autocalibrating parallel {MRI}: where {SENSE} meets {GRAPPA},'' \emph{Magnetic resonance in medicine}, vol.~71, no.~3, pp. 990--1001, 2014.

\bibitem{wang2004SSIM}
Z.~Wang, A.~C. Bovik, H.~R. Sheikh, and E.~P. Simoncelli, ``Image quality assessment: from error visibility to structural similarity,'' \emph{IEEE transactions on image processing}, vol.~13, no.~4, pp. 600--612, 2004.

\bibitem{ravishankar2010HFEN}
S.~Ravishankar and Y.~Bresler, ``{MR} image reconstruction from highly undersampled k-space data by dictionary learning,'' \emph{IEEE transactions on medical imaging}, vol.~30, no.~5, pp. 1028--1041, 2010.

\bibitem{otazo2015LS}
R.~Otazo, E.~Candes, and D.~K. Sodickson, ``Low-rank plus sparse matrix decomposition for accelerated dynamic {MRI} with separation of background and dynamic components,'' \emph{Magnetic resonance in medicine}, vol.~73, no.~3, pp. 1125--1136, 2015.

\bibitem{aggarwal2018modl}
H.~K. Aggarwal, M.~P. Mani, and M.~Jacob, ``Modl: Model-based deep learning architecture for inverse problems,'' \emph{IEEE transactions on medical imaging}, vol.~38, no.~2, pp. 394--405, 2018.

\bibitem{sandino2021DLESPIRIT}
C.~M. Sandino, P.~Lai, S.~S. Vasanawala, and J.~Y. Cheng, ``Accelerating cardiac cine {MRI} using a deep learning-based espirit reconstruction,'' \emph{Magnetic Resonance in Medicine}, vol.~85, no.~1, pp. 152--167, 2021.

\bibitem{huang2021LSNET}
W.~Huang, Z.~Ke, Z.-X. Cui, J.~Cheng, Z.~Qiu, S.~Jia, L.~Ying, Y.~Zhu, and D.~Liang, ``Deep low-rank plus sparse network for dynamic {MR} imaging,'' \emph{Medical Image Analysis}, vol.~73, p. 102190, 2021.

\bibitem{pan2024MCMR}
J.~Pan, M.~Hamdi, W.~Huang, K.~Hammernik, T.~Kuestner, and D.~Rueckert, ``Unrolled and rapid motion-compensated reconstruction for cardiac cine {MRI},'' \emph{Medical Image Analysis}, vol.~91, p. 103017, 2024.

\end{thebibliography}

\end{document}